\documentclass[review,authoryear]{elsarticle}

\usepackage{graphicx,psfig,pifont}
\usepackage[margin=1.5cm]{geometry}
\usepackage{natbib}
\usepackage{epsfig}
\usepackage{subfig}
\usepackage{enumitem}
\usepackage{multirow}
\newcommand{\bea}{\begin{eqnarray}}  \newcommand{\eea}{\end{eqnarray}}

\setitemize{leftmargin=*, nolistsep}
\usepackage[fleqn]{amsmath}
\usepackage{url}
\usepackage{xcolor}

\usepackage{lineno}

\journal{Renewable Energy}

\begin{document}

\begin{frontmatter}

\title{Variability of wave power production of the M4 machine at two energetic open ocean locations: off Albany, Western Australia and at EMEC, Orkney, UK}

\author[UWA,MERA]{J. Orszaghova}        \ead{jana.orszaghova@uwa.edu.au}
\author[ECN]{S. Lemoine}                \ead{siane.lemoine@gmail.com}
\author[TCOMS]{H. Santo}                \ead{harrif\_santo@tcoms.sg}
\author[UWA,MERA]{P. H. Taylor}         \ead{paul.taylor@uwa.edu.au}
\author[UWA,MERA]{A. Kurniawan}         \ead{adi.kurniawan@uwa.edu.au}
\author[UWA,MERA]{N. McGrath}           \ead{nicholas.mcgrath@uwa.edu.au}
\author[UWA]{W. Zhao}                   \ead{wenhua.zhao@uwa.edu.au}
\author[UWA,MERA]{M. V. W. Cuttler}     \ead{michael.cuttler@uwa.edu.au}

\address[UWA]{Oceans Graduate School, University of Western Australia, Crawley, WA 6009, Australia}
\address[MERA]{Marine Energy Research Australia (MERA), University of Western Australia, Australia}
\address[ECN]{Ecole Centrale de Nantes, 1 rue de la Noe, 44321 Nantes Cedex 3, France}
\address[TCOMS]{Technology Centre for Offshore and Marine, Singapore (TCOMS), Singapore 118411, Singapore}

\begin{abstract}
Since intermittent and highly variable power supply is undesirable, quantifying power yield fluctuations of wave energy converters (WECs) aids with assessment of potential deployment sites. This paper presents analysis of 3-hourly, monthly, seasonal, and inter-annual variability of power output of the M4 WEC. We compare expected performance from deployment at two wave energy hotspots: off Albany on the south-western coast of Australia and off the European Marine Energy Centre (EMEC) at Orkney, UK. We use multi-decadal wave hindcast data to predict the power that would have been generated by M4 WEC machines. The M4 machine, as a floating articulated device which extracts energy from flexing motion about a hinge, is sized according to a characteristic wavelength of the local wave climate. Using probability distributions, production duration curves, and coefficients of variation we demonstrate larger variability of the 3-hourly power yield at Orkney compared to Albany. At longer timescales, seasonal trends are highlighted through average monthly power values. From a continuity of supply perspective, we investigate occurrences of low production at three different threshold levels and calculate duration and likelihood of such events. Orkney is found to suffer from more persistent lows, causing a more intermittent power output. We also consider the effect of machine size on its power performance. Smaller machines are found to more effectively smooth out the stochastic nature of the underlying wave resource.
\end{abstract}


\end{frontmatter}


\section{Introduction}
\label{sec_intro}

The design of any wave energy converter (WEC) relies on knowledge of wave conditions at the intended deployment location. This is to ensure optimal performance (stemming from WECs' finite operational frequency-bandwidth), and to suitably minimise the risk of failure in severe conditions over the predicted lifespan of the device. Wave resource assessment is thus carried out to quantify and characterise the wave climate, based on measured or simulated wave data. The importance of considering the variability of wave climate at a particular site has been recognised for some time, but often wave resource assessments have focused on quantifying the magnitude of the raw resource via averaged quantities while neglecting temporal fluctuations. In reality, of course, the incident wave conditions vary over a wide range of time scales. This has implications on the power yield of WECs, affecting their intermittency and efficiency, which have not been extensively studied. In this paper, we investigate short-term, seasonal and inter-annual (i.e. year-on-year) variations in power production of the M4 WEC at two wave energy hot spots: off Albany in Western Australia and off Orkney, UK. This work builds directly upon \cite{santo2020_M4_EMEC_Albany}, who studied the mean power output performance of M4, as well as the device motions in survival mode, at the two locations.

\subsection{Wave resource and its temporal variability}
The incident wave conditions at any location are highly variable: fluctuations occur within an individual wave cycle, as well as on wave group, sea-state and storm scales, all the way up to seasonal and multi-year changes. The characteristics of the local wave climate may be derived from wave measurements (in-situ sensors such as wave buoys or remote sensors such as a wave radar), or from numerical hindcast models which can re-create multi-year records of wave conditions from the past. The available timeseries wave data is commonly synthesised into scatter diagrams, which provide sea-state occurrence probability distributions. In these, a sea-state is represented by a (wave height, wave period) variable pair, with no spectral shape and directional information. Alternatively, the local wave climate can also be described by the mean wave power density (typically expressed in kW/m). The annual wave power values allow for inter-annual trends to be studied, while mean monthly values can capture seasonal effects. 

There has been a multitude of wave power resource assessment studies covering many offshore and coastal domains, see for example \cite{coe2021} who provide tens of suitable references. Works of \cite{cornett2008}, \cite{gunn2012}, \cite{arinaga2012} and \cite{requero2015} consider the global oceans. Their analyses identify our two sites as rather energetic while also noting lower seasonal variation and thus a more stable resource along the southern Australian coast compared to western Europe. 

A number of more detailed, regional assessments have been carried out at both locations. The European Marine Energy Centre (EMEC) site, off Orkney, UK (see https://www.emec.org.uk/) has been included in the analyses of \cite{neill2013_NW_europe} and  \cite{neill2014_Orkney}, who report over five times more incident wave power in winter compared to summer. \cite{santo2015decadal} and \cite{santo2016extreme} consider longer time scales and provide decadal variability of the mean annual wave power and of the extreme wave conditions respectively for a number of locations in the North-East Atlantic, including Orkney. \cite{hughes2010} and \cite{hemer2017} quantify the national Australian wave energy resource. \cite{cuttler2020_seasonalWA} carried out a high spatio-temporal resolution study of the wave conditions off Albany in Western Australia. Here the seasonal variations of the wave energy resource are found to be much smaller, with winter around 2.5 times more energetic than summer.

\subsection{WEC performance and its temporal variability} 
Understandably, consistent wave conditions at a potential WEC deployment site are desirable. As noted by \cite{coe2021} and \cite{ringwood2015_EWTEC} for example, lower variability levels in the wave resource could reduce intermittency of power output and lead to higher capacity factors, as well as decrease the design demands associated with withstanding high loads. A WEC can be viewed simplistically as a band-pass filter acting on the incident wave energy flux; operating efficiently over a band of frequencies with a compromised performance elsewhere (see for example the M4 capture width ratio in Figure \ref{fig_RAO_and_CWR}). This suggests that the power production is likely to be smoother than the available incident wave power, as the resource fluctuations from extremes (i.e. large long waves) would be partially filtered out. In this work, we confirm this to be the case for both the Albany and Orkney locations.   

A number of studies in the literature have assessed consistency of the power supply of wave energy converters. \cite{carballo2015} consider intra-annual/seasonal trends of five different wave energy technologies at two locations off northern Spain. They find between 1.5 to almost 4 times more production during the most energetic winter months, compared to the quiescent summer months. However, it appears that in their study the characteristics of the devices have not been matched to the local wave climates. 
\cite{morim2019} analyse both intra- and inter-annual (year-on-year) variability of power production of ten different WECs off New South Wales along the south-eastern Australian coast. They highlight the importance of optimal sizing of devices. Their results indicate much lower variability in the power production, across both timescales, for correctly sized WECs compared to oversized non-optimal ones. They also note that the magnitude of the inter-annual variations in the power production (for suitably downsized WECs) is generally lower than the inter-annual variability in the wave resource.

\cite{penalba2018} and \cite{ulazia2019} consider decadal changes in WEC performance in the North-East Atlantic, off the western European coast. They note a progressive increase in the incident wave energy flux and the mean annual WEC power yield, respectively up to 40\% and 30\% higher values during 1980-2000 compared to 1900-1920. When the survival behaviour of the WECs is taken into account (i.e. no production during too severe conditions), the long-term upwards trend in absorbed power is still notable, though somewhat reduced, while the WEC efficiency (ratio of absorbed to available incident power) actually declines. Both of these are due to increasing occurrence of severe conditions during the 20\textsuperscript{th} century. \cite{santo2016decadalM4} study the historical power production potential of an M4 WEC at Orkney, dating back over more than 300 years. The reconstructed time series reveal $\pm 20 \%$ fluctuations in annual absorbed power values, which are much smaller than that of the wave power resource. In contrast to the western Europe locations, a study for the Chilean coast (see \cite{ulazia2018}) found the wave resource and the WEC power output to be remarkably stable over the studied 100 years.  







\subsection{M4 wave energy converter}
\begin{figure}[ht]
    \centering
    {\includegraphics*[viewport= 142 378 468 479, width=0.49\linewidth]{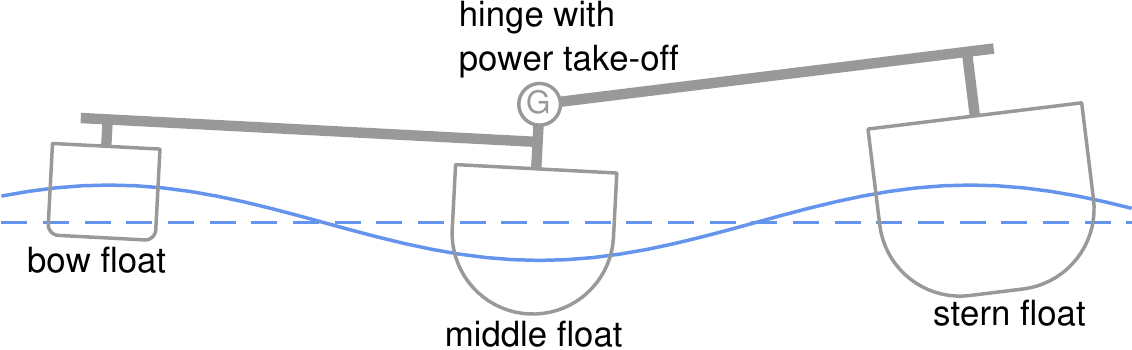}}
    \caption{Schematic diagram of the three-float M4 WEC.}
    \label{fig_M4}
\end{figure}

A wide variety of different wave energy converters designs have been proposed (see \cite{falcao2010} and \cite{babarit2017} for example). In this study, we use the M4 WEC, which falls into the category of wave-activated devices, meaning that the device, or parts of it, are excited by wave action and the rigid body (or elastic) response drives the power take-off machinery. The original M4 design consists of three collinear surface-piercing floats, each float with a circular cross-section in plan (see Figure \ref{fig_M4}). The front two floats are rigidly connected via a beam above the free surface. The rear float is rigidly connected to a second beam, which terminates at a hinge point above the mid float. Power is generated due to relative angular motion between the two bodies. The whole system is soft moored and the progressively increasing float diameters aid weathervaning such that the device self-aligns to the wave propagation direction. The device has been extensively studied, with over 15 peer-reviewed journal publications to date. The M4 development pathway has focused on innovation and design optimisation via numerical simulations and experimentally at a reduced scale, thus achieving high Technology Performance Level (TPL) before advancing its Technology Readiness Level (TRL) via larger-scale ocean deployments. Such an approach has been suggested by \cite{weber2012_TPL} for example. The early studies (e.g. \cite{stansby2015_RenEnrg}, \cite{stansby2015_JOEME}) documented the working principles of the WEC, highlighting its broadbanded and relatively high capture width. A number of numerical models for the device dynamics, based on linear potential flow theory, have been created (e.g. \cite{eatock_taylor2016}, \cite{sun2016_JOEME}, \cite{stansby2016_model}) and have been used to investigate effects of geometric variations (such as float separation distances and hinge elevation) on the device power absorption. Further optimisation was carried out by \cite{stansby2017_multifloat} who considered multiple mid and stern floats to increase the techno-economic performance of M4. Recent advances focused on real-time control (see for example \cite{liao2020_M4control} and \cite{zhang2021_M4control}), while survivability has been investigated experimentally and numerically by \cite{santo2017_extreme_M4_JFM} and \cite{santo2020_M4_EMEC_Albany}.


\subsection{Aims, novelty and structure of this paper}
The wealth of existing studies, including publicly available performance data, makes M4 an ideal technology to adopt for the investigation of absorbed power variability pursued herein. We build on the work of \cite{santo2020_M4_EMEC_Albany} who characterised the performance of M4 at Albany, Western Australia and at EMEC, Orkney, UK. In this study, however, for Albany we use 38 years long hindcast wave data from \cite{cuttler2020_seasonalWA} instead of the 8 years long wave buoy record used by \cite{santo2020_M4_EMEC_Albany}. Compared to \cite{santo2016decadalM4} who analysed decadal variability of M4 power output at Orkney, we consider fluctuations over shorter time scales, spanning 3-hourly, seasonal and inter-annual changes. 

There is a need to understand temporal variability of WEC power production across different time scales. Although it arises from the resource, the variability is strongly dependent on the device's operational range. Absorbed power fluctuations have direct implications on the cost of wave energy harvesting, which is currently considered too high to be cost-competitive with other energy sources (\cite{babarit2017}). It is beyond the scope of this paper to carry out a detailed cost analysis. However, quantifying the variability of the power production of the M4 WEC provides new insight on the potential of wave energy, which goes beyond long-term averaged performance indicators. Importantly, we investigate the role of device sizing when assessing WEC power variability. We also attempt to draw comparisons to wind energy from published literature. 

The paper is structured as follows. We first introduce the available wave hindcast datasets and explain the methodology for calculation of the absorbed power. The calculated M4 power data time series are then processed statistically, with graphically presented results to compare and contrast the two locations of interest. We examine distributions of 3-hourly absorbed power, and investigate in detail occurrences of minimal power output characterising their frequency and persistence. We also assess monthly trends at the two locations, and consider the effect of device sizing on seasonal variability and intermittency of the power yield.
\begin{figure}[ht]
    \centering
    {\includegraphics*[viewport= 157 325 419 514, width=0.4\linewidth]{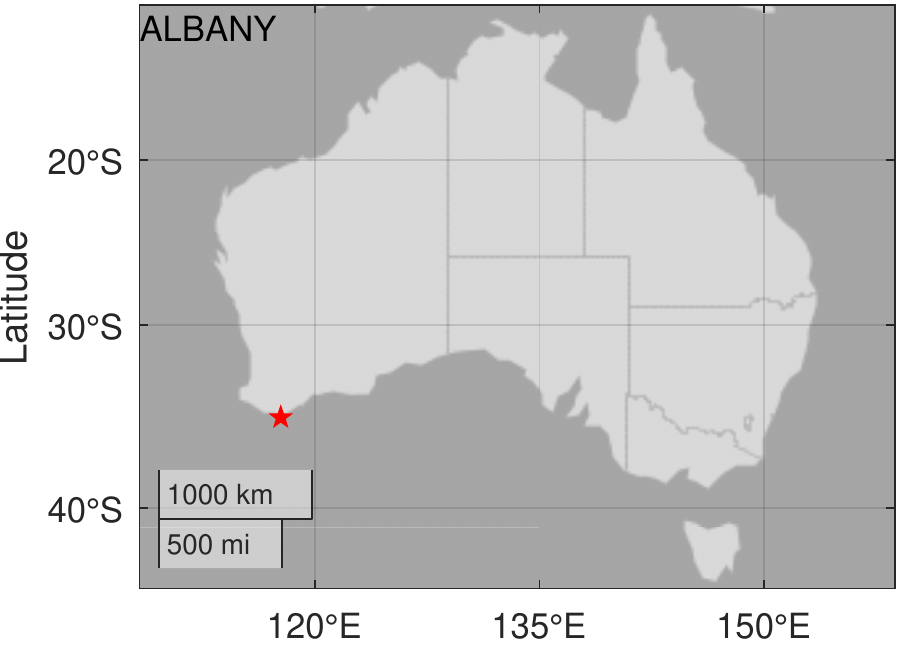}} \qquad \qquad \qquad 
    {\includegraphics*[viewport= 157 325 419 514, width=0.4\linewidth]{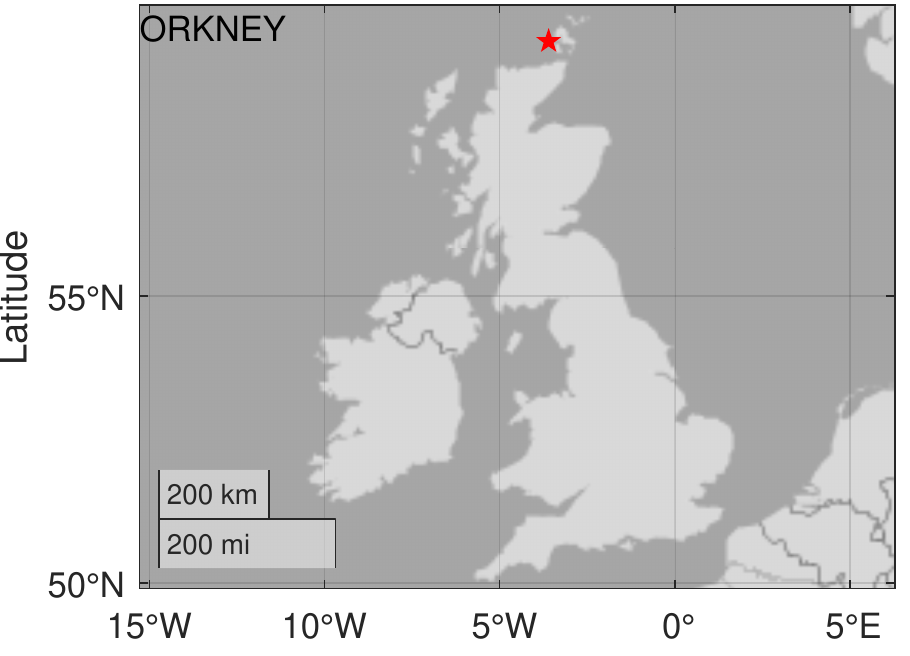}}
    {\includegraphics*[viewport= 157 324 419 508, width=0.4\linewidth]{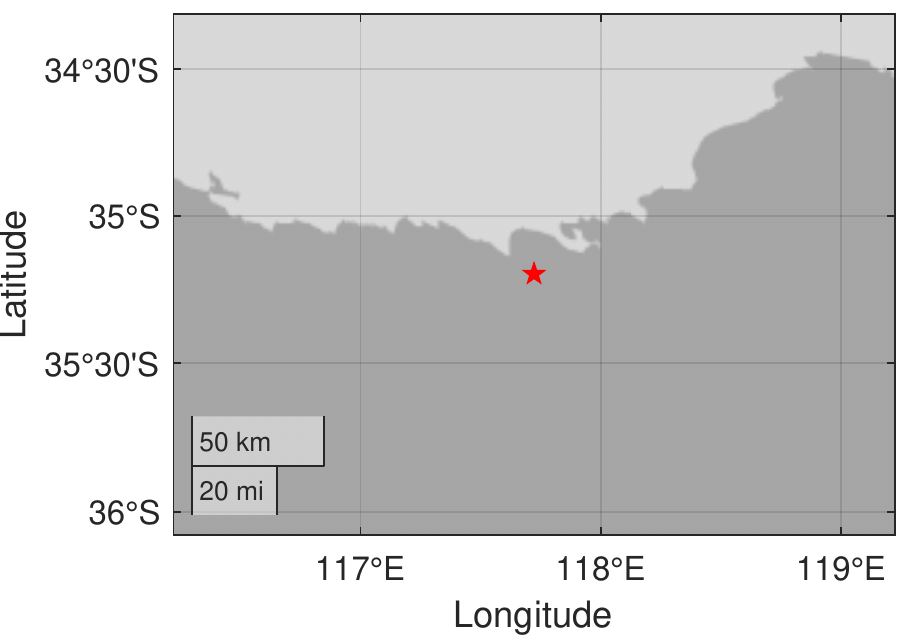}} \qquad \qquad \qquad
    {\includegraphics*[viewport= 157 324 419 508, width=0.4\linewidth]{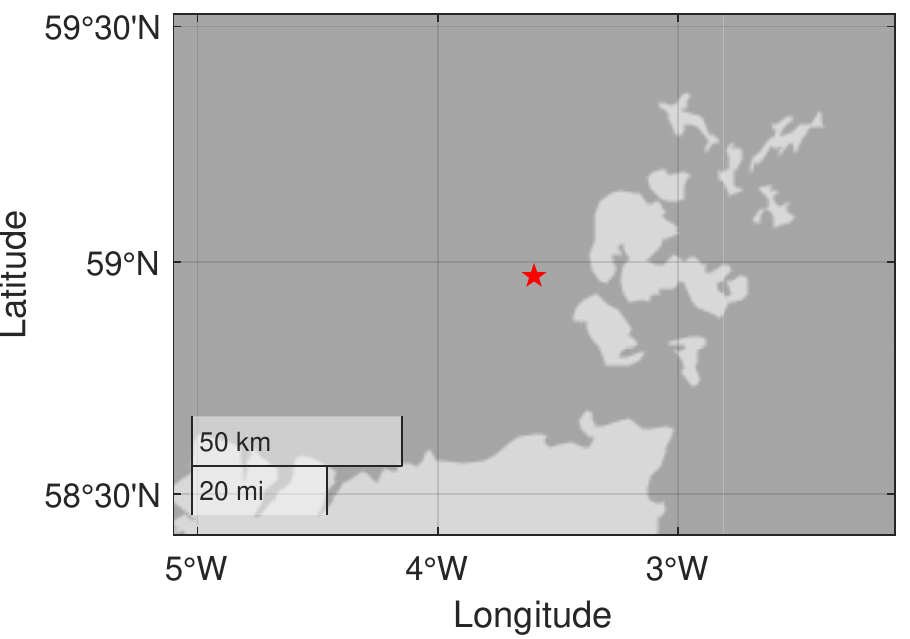}}
    \caption{Left: Map of Australia with the Albany site on the south-west coast of Western Australia highlighted (zoomed-in view bottom-left). Right: Map of the UK with the EMEC site, west of Orkney highlighted (zoomed-in view bottom-right). Maps have been generated with an in-built function geoplot in \textsc{Matlab}.}
    \label{fig_map}
\end{figure}

\section{Data and methods}
\label{sec_data}

We use hindcast wave data for both locations. For Albany, 38 years of wave data are sourced from \cite{cuttler2020_seasonalWA}. For EMEC in the Orkneys, Scotland, 54 years of wave data come from \cite{santo2016decadalM4}. This section describes the available datasets and outlines the incident wave power and the absorbed power calculations.

The Albany wave data is from a high spatio-temporal hindcast over $1980 - 2017$, with three nested grids with resolutions of 0.5, 0.165 and 0.05 km. The site of interest is located at S $35^{\circ} \ 11^{\prime} \ 52.8^{\prime\prime}$, E $117^{\circ} \ 43^{\prime} \ 19.2^{\prime\prime}$, approximately 15 km offshore, in 60 m water depth, and is fully exposed to waves from the south and south west approaching across the Southern Ocean (see Figure \ref{fig_map}). The chosen location coincides with a Datawell Directional Waverider Mark III wave buoy, operated by the Western Australia Department of Transport, which had been used for validation of the hindcast. The wave data is hourly, in the form of two-dimensional free-surface variance density spectra. However, for consistency with the Orkney data, in our calculations we utilise one-dimensional JONSWAP spectral shapes and consider 3-hourly intervals (by simple averaging of the available hourly Albany data). We note that the long-crested assumption is justified as the directional spread is rather narrow; over 50\% of sea-states annually exhibit (standard deviation of) directional spreading below $10^\circ$ (see \cite{hlopheTBC_multidir_AOR}). More details on the calculation of the spectra is provided in the next paragraph and in the Appendix. The Albany hindcast dataset is continuous (with no data gaps) and almost 5 times longer than the 8-year long wave buoy record used by \cite{santo2020_M4_EMEC_Albany}, both of which are advantageous for the analysis herein.

The Orkney wave data is from the Norwegian 10 km Reanalysis Archive (NORA10) hindcast from $1958 - 2011$ (see \cite{reistad2011high}). The location of the NORA10 grid point is N $58^{\circ} \ 58^{\prime} \ 12^{\prime\prime}$, W $3^{\circ} \ 36^{\prime} \ 0^{\prime\prime}$, which is approximately 15 km west of the EMEC test site for marine renewable energy machines on the west coast of Orkney (see Figure \ref{fig_map}). The water depth is assumed to be 60 m. The NORA10 wave data available is in 3-hour intervals, with bulk parameters of significant wave height $H_s$, peak spectral wave period $T_p$, mean wave period $T_m$, wind speed, wind and wave directions. We use JONSWAP spectra calculated using the hindcast bulk parameters $H_s$ and $T_p$, with the value of the peak enhancement parameter $\gamma$ chosen so as to match the hindcast $T_m$. Since spectral bandwidth plays an important role in WEC power output calculations (see \cite{Saulnier2011}), our procedure helps achieve reasonably representative energy distribution (across frequencies) in the absence of the full spectral content. More information on this fitting method is provided in the Appendix A, together with indication of the appropriateness of this approximation. 

To quantify the wave resource, we calculate the incident wave power per unit length of wave-front for each sea-state (also known as the wave energy flux density or the wave power density) according to  
\begin{eqnarray}
\label{eq_Pinc_finite}
P_{wave} = \rho g \int S(f) c_g(f) \ \mathrm{d} f \ ,
\end{eqnarray}
where $\rho$ is the water density, $g$ is the gravitational acceleration, $S(f)$ and $c_g(f)$ are the wave spectrum and the group velocity, both of which vary with frequency $f$. The above formulation is suitable for any water depth $h$, with the group velocity given by $c_g = \frac{1}{2}(1 + \frac{2 k h}{\sinh(2 k h)}) \frac{\omega}{k}$. The angular wave frequency $\omega = 2 \pi f$ (or wave period $T=\frac{1}{f}$) and the wavenumber $k$ (or wavelength $\lambda=\frac{2 \pi}{k}$) are related through the linear dispersion relation
\begin{eqnarray}
\label{eq_disp}
\omega^2 = g k \tanh(k h) \ .
\end{eqnarray}


\begin{figure}[ht]
    \centering
    {\includegraphics*[viewport= 128 307 450 518, width=0.49\linewidth]{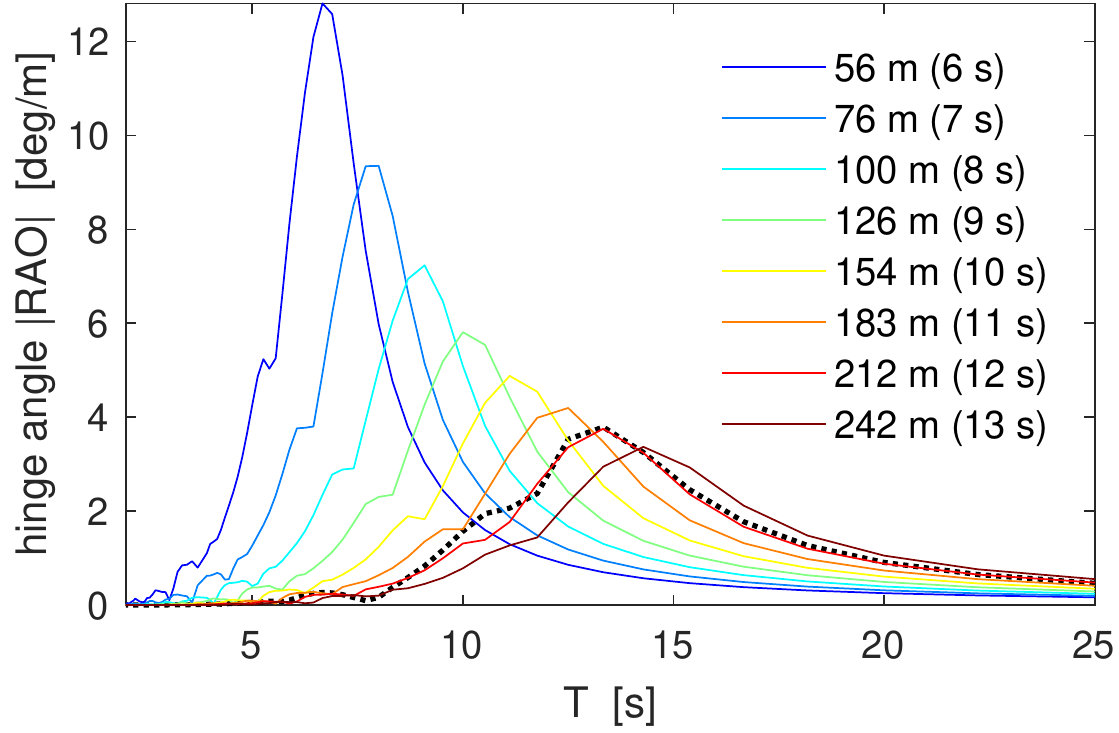}}
    {\includegraphics*[viewport= 128 307 450 518, width=0.49\linewidth]{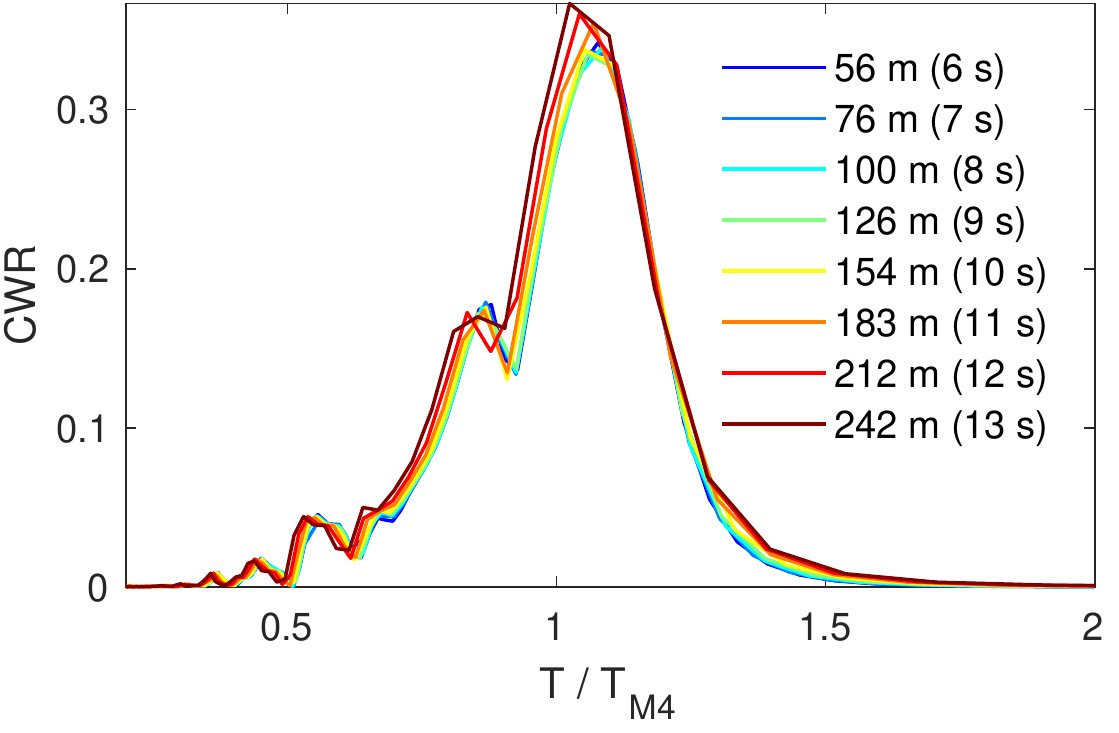}}
    \caption{Left: Hinge angle per unit amplitude wave as a function of wave period $T$, i.e. modulus of the hinge angle response amplitude operator (RAO), for different machines sizes. For comparison, the black dotted line represents the RAO curve from the numerical model of \cite{eatock_taylor2016}. Right: Capture width ratio $CWR$ as a function of non-dimensional wave period (incident wave period $T$ divided by the period $T_{M4}$ which corresponds to the wavelength of the machine length). Note that the legend gives the M4 machine lengths, with the corresponding wave periods given in brackets. All for 60 m water depth.}
    \label{fig_RAO_and_CWR}
\end{figure}

The practical wave power available for conversion at each location can be obtained by incorporating the capture width characteristics of the M4 wave energy converter into the wave power calculation. The capture width quantifies the WEC performance, and is defined as the ratio between the absorbed power and the incident wave power available per unit crest-length. It is a length-scale corresponding to the width of wave crest conveying the absorbed power. The capture width ratio $CWR$ is a non-dimensional quantity, which quantifies the WEC hydrodynamic absorption efficiency. It is a frequency-dependent quantity, and here it is given as the capture width divided by the corresponding wavelength $\lambda(f)$. Our non-dimensionalisation follows from the fact that, for the M4 WEC, the characteristic length-scale is along the wave propagation direction, and the device is sized such that the bow to stern floats are a typical wavelength apart to maximise the hinge rotation. Conventionally, the width of the machine is used for the non-dimensionalisation (see Section 4.1 in \cite{babarit2017}). We use a frequency-domain hydrodynamic model to calculate the motion response and power absorption of the M4 WEC. The linear hydrodynamic coefficients are evaluated by the radiation-diffraction software HydroStar (see \cite{hydrostar}). To represent the rotational motion at the hinge, we utilise generalised modes (see \cite{Newman1994_gen_modes}). The device geometry is close to that depicted in Figure \ref{fig_M4}, as also used by \cite{santo2016decadalM4} and \cite{santo2020_M4_EMEC_Albany}. The same (Froude-scaled) power take-off damping coefficient as in these studies is applied, and the device is assumed to not be actively controlled. The calculated hinge angle response, i.e. the power-producing mode of motion, is shown in Figure \ref{fig_RAO_and_CWR} for a range of machine sizes (all in 60 m depth). For reference, the equivalent response amplitude operator (RAO) curve, appropriately Froude scaled for 60 m depth, from the hydrodynamic model of \cite{eatock_taylor2016} is also plotted, as a check on our calculations. Also shown in Figure \ref{fig_RAO_and_CWR} is the capture width ratio, which we use for the absorbed power calculations as outlined below. As expected, the non-dimensional $CWR$ curves are close to identical, especially for the smaller device sizes, indicating that the different devices' performance is self-similar and effectively independent of the water depth. We recall that we do not take directional spreading into account, and assume the waves to be uni-directional.

Following \cite{santo2016decadalM4} and \cite{santo2020_M4_EMEC_Albany}, the methodology to obtain the absorbed power on a sea-state basis, i.e. the mean power in a 3-hourly interval, is as follows:
\begin{enumerate}
\item Size the M4 machine based on the wavelength corresponding to the long-term average energy period $T_e$ of the local wave climate. For Albany, we calculate the energy period using the hindcast wave spectra according to $T_e = m_{-1} /m_0$, where $m_n = \int f^n S(f) \ \mathrm{d} f$ is the $n^{\mathrm{th}}$ spectral moment. For Orkney, in the absence of hindcast spectra, we estimate the energy period according to $T_e = (T_p + T_m)/2$, which was shown to be a suitable approximation by \cite{santo2016decadalM4}). Note that in Section \ref{sec_WEC_size} other device sizes will also be considered. 
\item For each sea-state, compute the mean 3-hourly absorbed power $P$, using the incident wave energy flux $P_{wave}$ formulation from Equation \ref{eq_Pinc_finite}, and incorporating the capture width ratio and assumed 90\% mechanical efficiency as per
\begin{eqnarray}
\label{eq_P}
P &=& 0.9 \ \rho g \int S(f) \ c_g(f) \ CWR(f) \ \lambda(f) \ \mathrm{d} f \ \\
  &=& 0.9 \ B \int S(f) \ \Big( 2 \pi f  \ |RAO(f)| \Big)^2 \ \mathrm{d} f \nonumber ,
\end{eqnarray}
where $B$ denotes the power take-off damping coefficient and the hinge angle RAO is measured in units of rad/m. We note that the two above formulations are equivalent. Analogous methods for evaluating WEC performance on a sea-state basis were used by  \cite{carballo2015} and \cite{morim2019} for example, who used sea-state power matrices from the literature or provided by the WEC developer, while \cite{penalba2018} and \cite{ulazia2019} for example calculated the absorbed power using the authors' own time-domain hydrodynamic models of the studied WECs. 
\item Obtain the mean absorbed power over whole duration of the available wave data, and assume the machine capacity to be $3 \times$ this long-term average, such that any 3-hourly power output cannot exceed the capacity and simply saturates at this limit. The device is assumed to continue operating but in such a way that any power in excess of the machine capacity is lost; the power is effectively clipped. 

Re-compute the practical power including clipping until convergence. We note that only a small number of iterations is needed for the results to converge. The final converged long-term mean absorbed power is denoted by $\bar{P}$. We note that the capacity ($3 \times \bar{P}$) is effectively the rated power of the machine. The capacity factor is thus set to $33\%$, which is comparable to modern onshore wind turbines (see \cite{Burton2021}).
\end{enumerate}

In addition to the standard sizes, we also consider a range of smaller and larger devices, to see the dependence of the absorbed power variations on the WEC size. The power calculation steps are the same as outlined above, apart from the first step, where the device length is simply set.

\section{Variability of M4 power output}

In this section, we describe and assess the variability and distribution of practical wave power produced by the M4 machine (with clipping) for the Albany and Orkney locations, which has been grouped into 3-hourly, monthly and seasonal intervals, over the 38-year and 54-year hindcast data sets. The properties of the machines sized for the respective wave climates are discussed first. The long-term average $T_e$ (throughout the entire length of the respective hindcast data sets) for Albany is 11.1 s and for Orkney 8.4 s. This corresponds to the overall machine length being 186 and 111 m respectively for the 2 locations. For reference, the float diameters (from bow to stern) would be approximately 23, 35 and 46 m at Albany, compared to 14, 21 and 28 m at Orkney. We note that from the available 8-year wave buoy data at Albany, collocated with the hindcast grid point, \cite{santo2020_M4_EMEC_Albany} evaluated the long-term average $T_e$ to be 10.5 s. This is consistent with \cite{cuttler2020_seasonalWA} who point out that the hindcast wave periods are slightly over-predicted with a bias of approximately 0.5 s. The long-term average practical power $\bar{P}$ at Albany is 1.5 MW and at Orkney 300 kW, with the machine capacities being 4.5 MW and 900 kW respectively. 

\subsection{3-hourly power distribution}
\begin{figure}[ht!]
\centering
{\includegraphics*[width=0.49\linewidth, viewport=120 303 447 520]{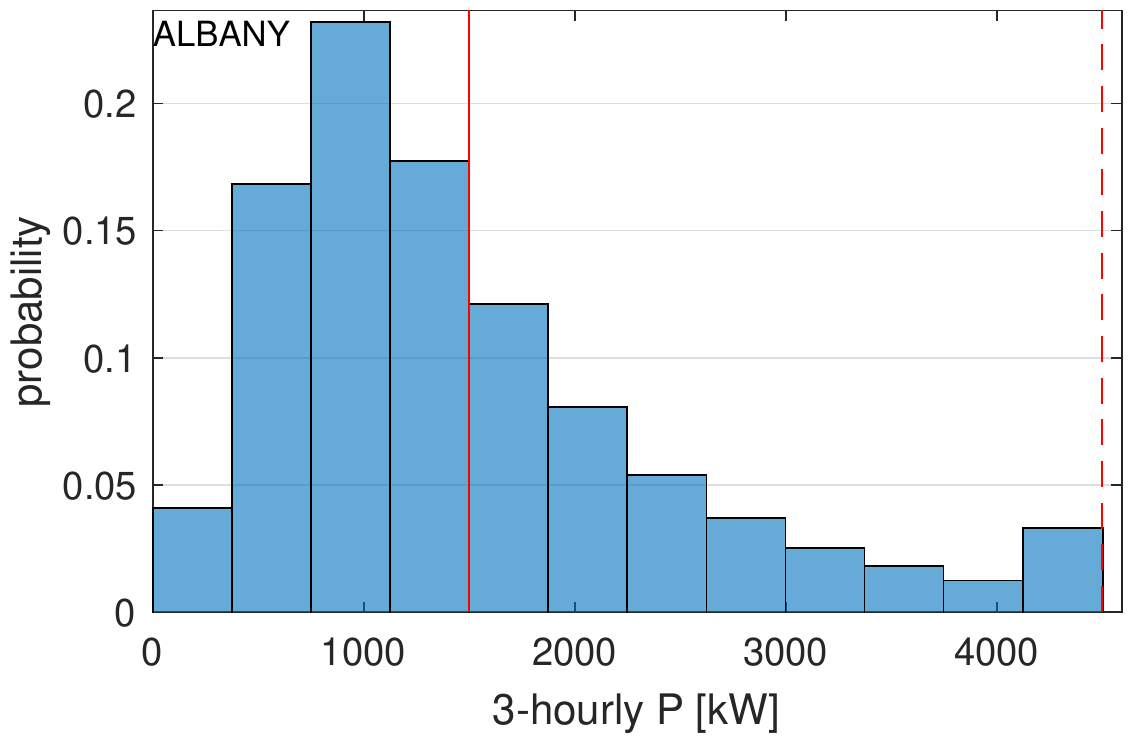}}
{\includegraphics*[width=0.49\linewidth, viewport=120 303 447 520]{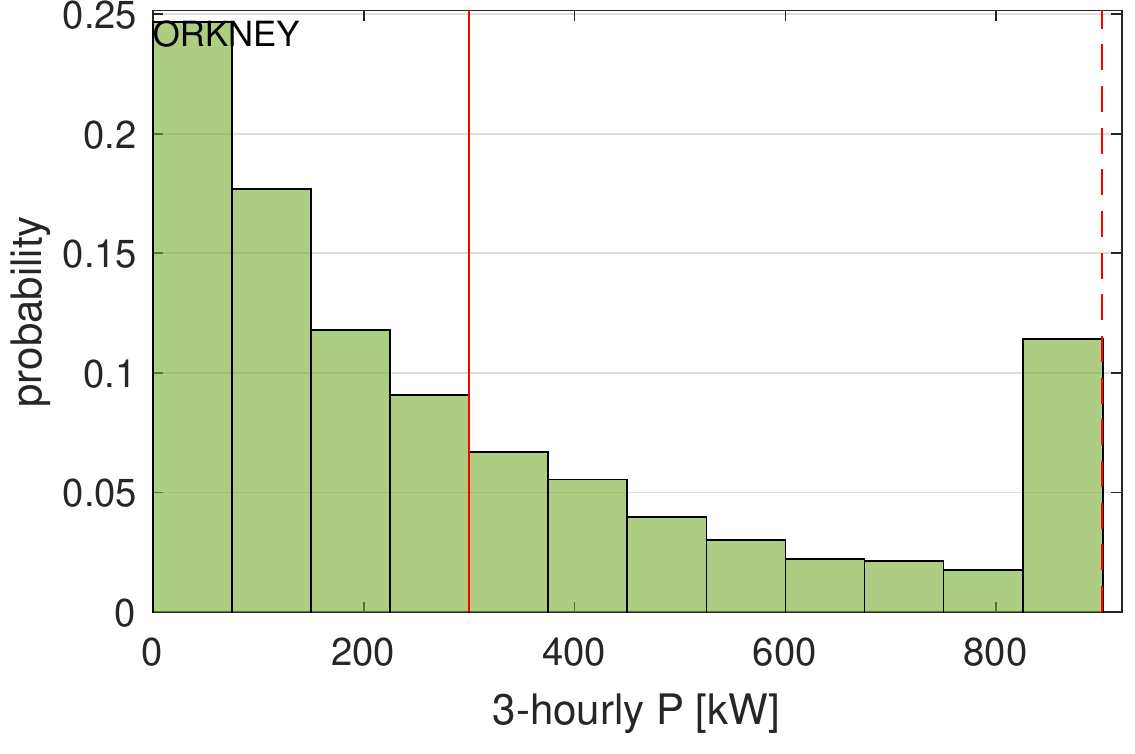}}
\caption{Histogram of average annual 3-hourly absorbed power $P$ (with clipping) for Albany on the left and for Orkney on the right. The long-term average practical power $\bar{P}$ is shown by the solid red line, while the dashed red line represents machine capacity. Note that the last bin contains clipping instances.}
\label{fig_histogram}
\end{figure}

Figure~\ref{fig_histogram} shows the histogram of 3-hourly power distribution produced by the M4 machine at Albany during $1980 - 2017$ and at Orkney during $1958-2011$. The power data (below the machine capacity) has been grouped into twelve equal bins, such that the width of each bin corresponds to $0.25 \bar{P}$. The height of each bar represents the probability of occurrence of the corresponding power levels throughout the entire records. The last bin represents power levels $2.75 \le P / \bar{P} \le 3$ and as such also contains instances of power clipping. The long-term average practical power $\bar{P}$ at each location is shown by the solid red line, while the machine capacity is denoted by the dashed red line. The distributions at the two locations are rather different, though both exhibit a positive skew. In agreement with observations made by \cite{coe2021}, the characteristic shape of the distribution, with a long high-power tail, follows from the resource. For reference, the corresponding histogram plots for the incident wave power can be found in Figure \ref{fig_histogram_Pwave} in Appendix B. The distribution tail arises from rare energetic storms. However, in Figure \ref{fig_histogram}, the absorbed power tail has effectively been clipped due to the finite machine capacity.  Overall, the power distribution is much more positively skewed at Orkney. We note in passing that the distribution at Orkney is similar to power generated from a wind turbine as per \cite{holttinen2005}. At Albany, the distribution peaks around 1 MW, which is roughly $2/3$ of the long-term mean machine power $\bar{P}$. At Orkney, the most common power outputs are a small fraction of the long-term mean, meaning that the machine produces rather low power levels ($P \le 0.25 \bar{P}$) approximately a quarter of the time on average. In contrast, such low power instances are six times less likely for the M4 machine at Albany. Frequent occurrences of minimal power yield are not desirable, as these correspond to a highly intermittent supply. From this perspective, Albany appears to be a more suitable location for M4 WEC deployment. Further analysis of low production events is presented in Section \ref{sec_low_P}.

\begin{figure}
\centering
{\includegraphics*[width=1\linewidth, viewport=22 281 540 570]{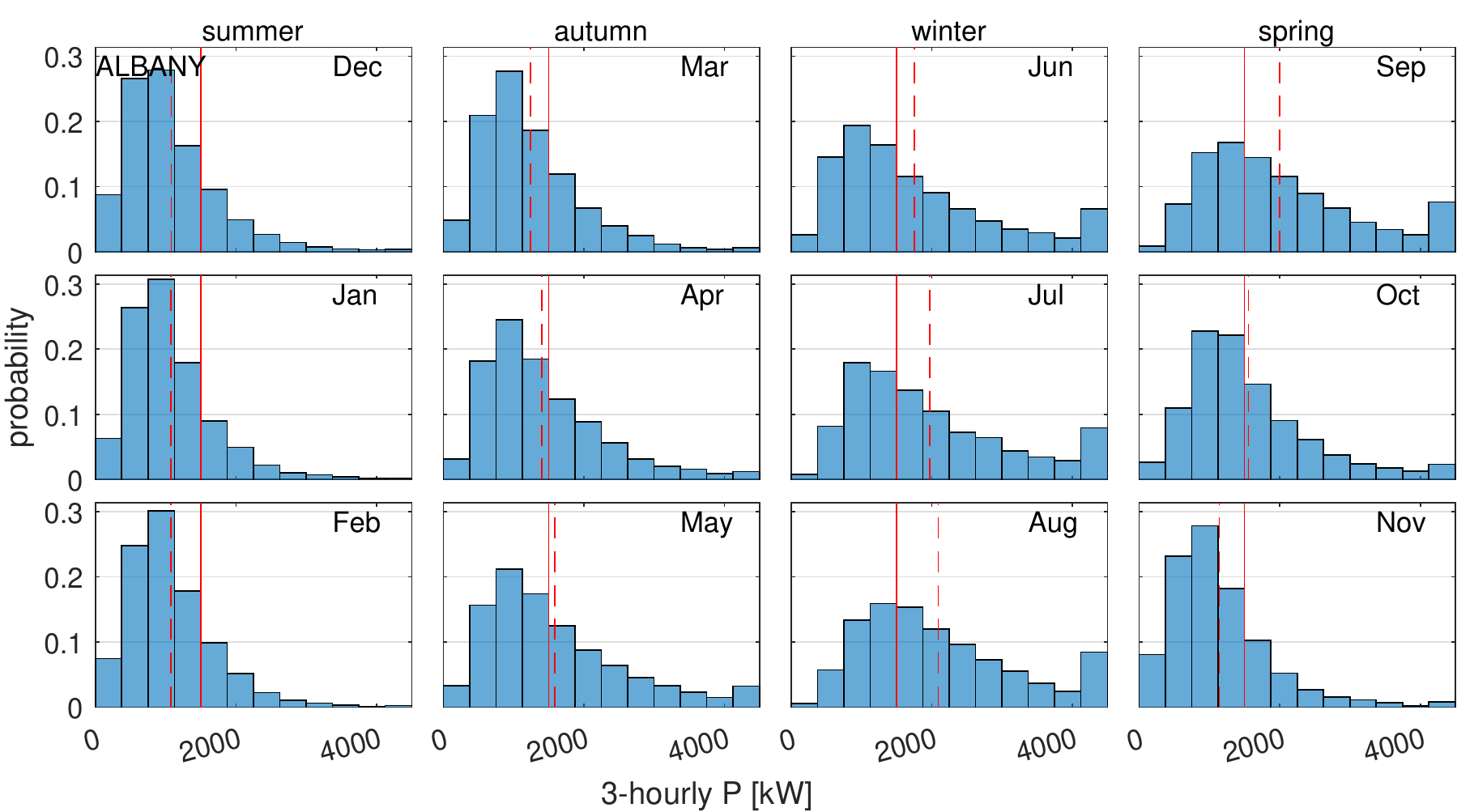}}
{\includegraphics*[width=1\linewidth, viewport=22 281 540 570]{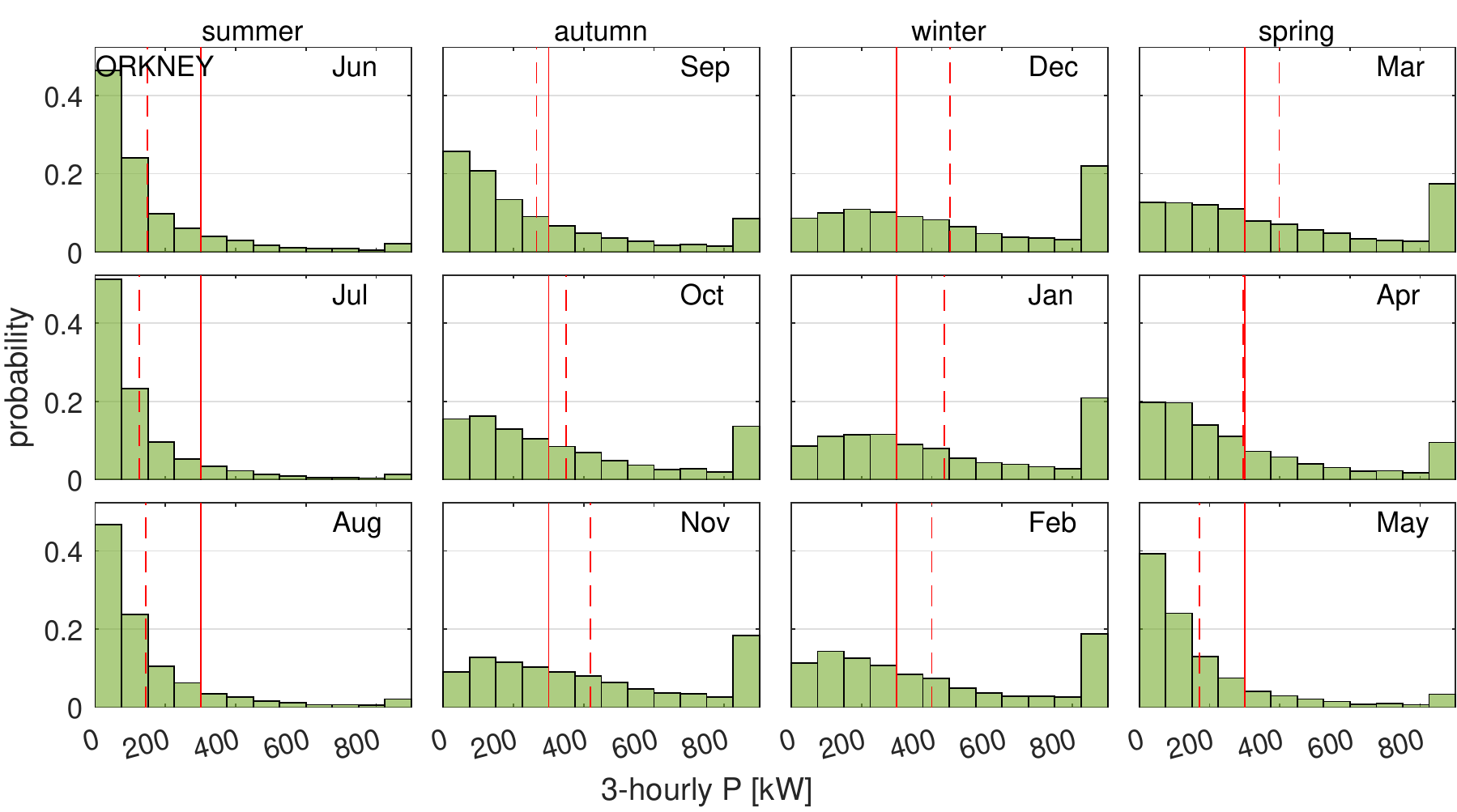}}
\caption{Histogram of 3-hourly absorbed power $P$ (with clipping) split into 12 months for Albany at the top and for Orkney at the bottom. The seasons at the two locations have been aligned, such that the columns (of subplots) from left to right correspond to summer, autumn, winter and spring. The long-term average practical power $\bar{P}$ is shown by the solid red line, while the dashed red line represents the long-term average for each month. Note that the last bin in each histogram contains clipping instances.}
\label{fig_histogram_monthly}
\end{figure}

Figure~\ref{fig_histogram_monthly} presents a closer look at the 3-hourly power distributions over the full hindcast data sets, which have now been split into twelve months. Each column of three sub-plots represents a season. Note that for easier comparison, the seasons are aligned (for example the first column corresponds to summer: December, January and February in the southern hemisphere and June, July and August in the northern hemisphere). As before, the long-term average practical power $\bar{P}$ is shown by the red solid line. Additionally, the long-term average for each month is shown by dash-dotted lines. Seasonal variation is evident at both locations. At Albany, during the austral summer (December, January, February), the distribution is more skewed to the left resulting in lower monthly mean power output due to a more benign wave environment, while power levels above $2 \times \bar{P}$ are very rare. On the other hand, in winter (June, July, August), the distribution is more evenly spread over the full range of power levels. The monthly averages are above the long-term mean $\bar{P}$ due to a rougher wave environment, and the machine power is clipped more frequently. Similar trends are observed at Orkney. We note, however, the much higher probability of low power levels below $0.25 \bar{P}$ at Orkney during summer (June, July, August), reaching over 45\% of the time during this season. We note a close correspondence between the absorbed power distributions and those of the resource (see Figure \ref{fig_histogram_Pwave_monthly} in Appendix B). This points to the fact that any findings on the deployment site suitability, based on our calculations for the M4 device, could be more universally extended to other WEC technologies.

\begin{figure}[ht]
\centering
{\includegraphics*[width=0.49\linewidth, viewport=127 308 479 522]{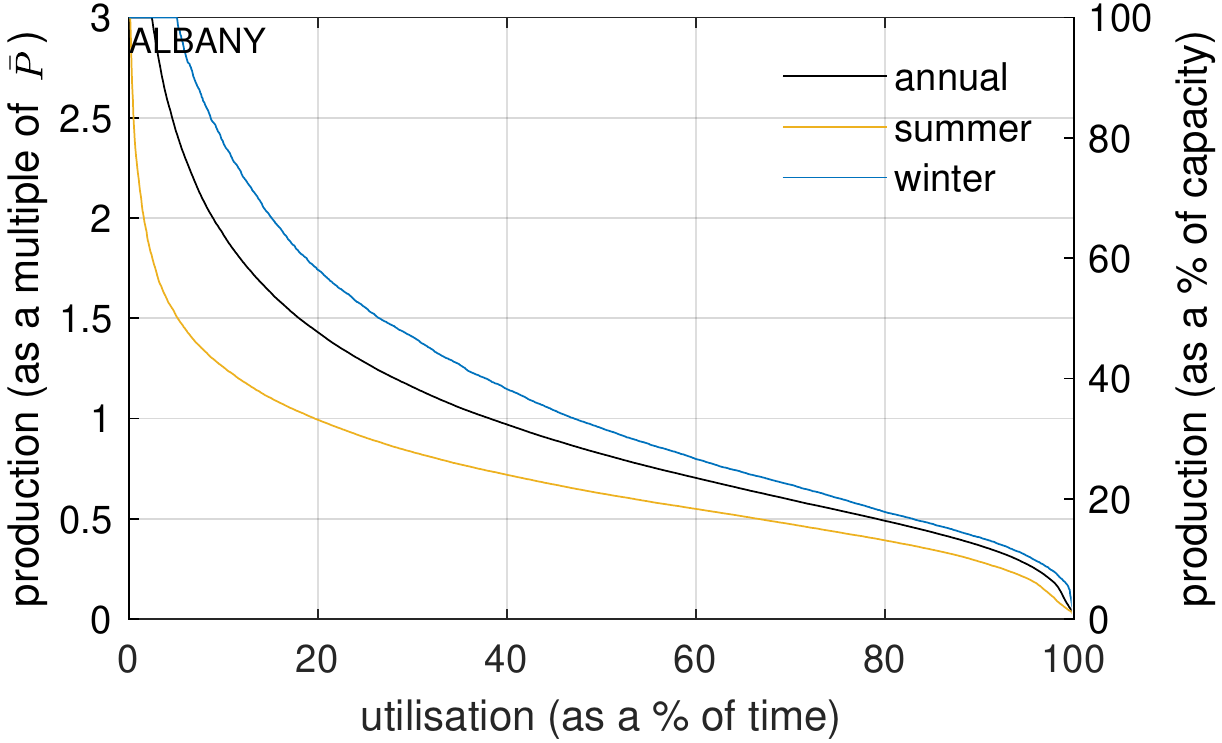}}
{\includegraphics*[width=0.49\linewidth, viewport=127 308 479 522]{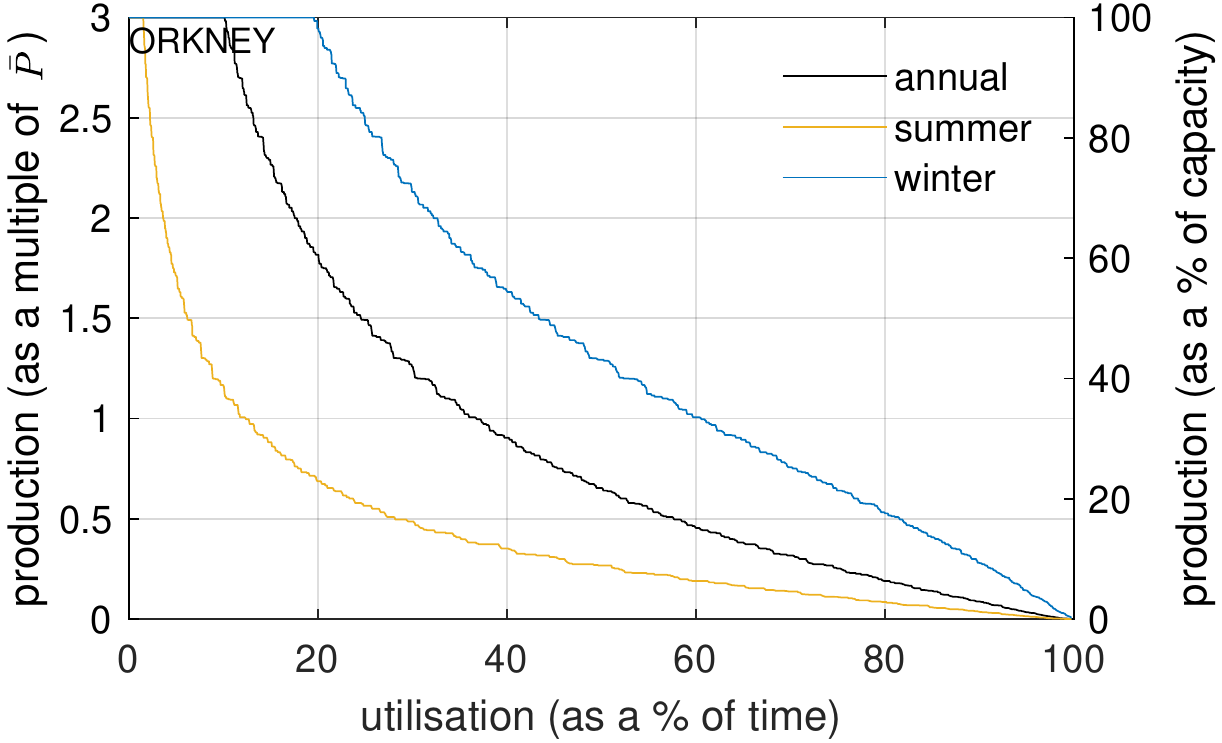}}
\caption{Production duration curves for the 3-hourly absorbed power $P$ (with clipping) for Albany on the left and for Orkney on the right. The black curve represents the duration curve from the entire hindcast. The yellow curve represents duration curve for summer (all December, January and February records for Albany; all June, July and August records for Orkney), while the blue curve represents duration curve for winter (all June, July and August records for Albany; all December, January and February records for Orkney).}
\label{fig_duration_curve}
\end{figure}

The strong seasonal variation observed above justifies splitting the total practical wave power into seasons. We do this in Figure \ref{fig_duration_curve}, which shows the production duration curves for the summer (yellow) and winter (blue) periods. The curves are created by sorting (in descending order) all 3-hourly absorbed power values within the summer and winter periods of the hindcast. Additionally, we also process the entire hindcast dataset in this way; the resulting curve (black) represents the mean annual trend. The curves are akin to a cumulative probability distribution, and express the proportion of time that the power output exceeds a particular value (see \cite{Renew_el_power_systems_book}). Flatter curves indicate more steady production; a horizontal line would correspond to a constant power output with no variability. At both locations the production is more steady, though lower, during summer compared to winter. Albany exhibits a more consistent power supply compared to Orkney, which is considered favourable. In Albany the power yield $P > 0.5 \bar{P}$ occurs close to 80\% of the time on average, and even during the summer season roughly 65\% of the time. On the other hand, in Orkney such power yields occur roughly 60\% of the time annually, while only 30\% during the summer season. We note in passing that duration curves have been used for example by \cite{holttinen2005} and \cite{katzenstein2010variability} for power generation assessment of wind turbines/farms.

\begin{figure}[ht]
\centering
{\includegraphics*[width=0.49\linewidth, viewport=121 315 446 521]{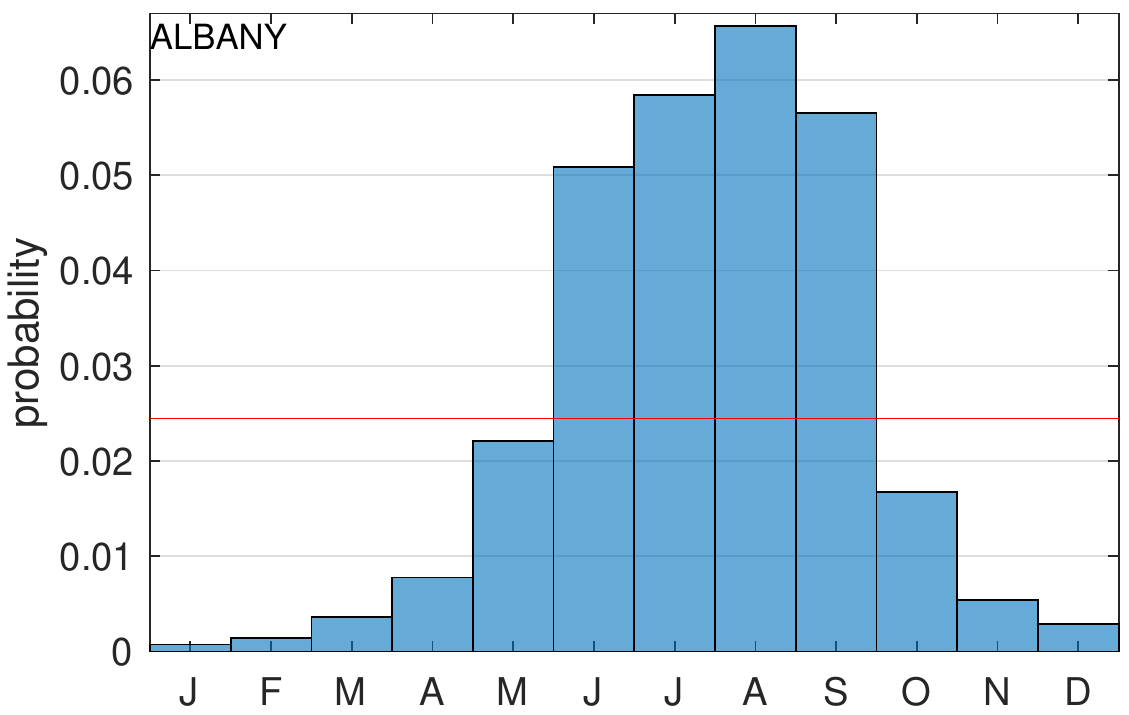}}
{\includegraphics*[width=0.49\linewidth, viewport=121 315 446 521]{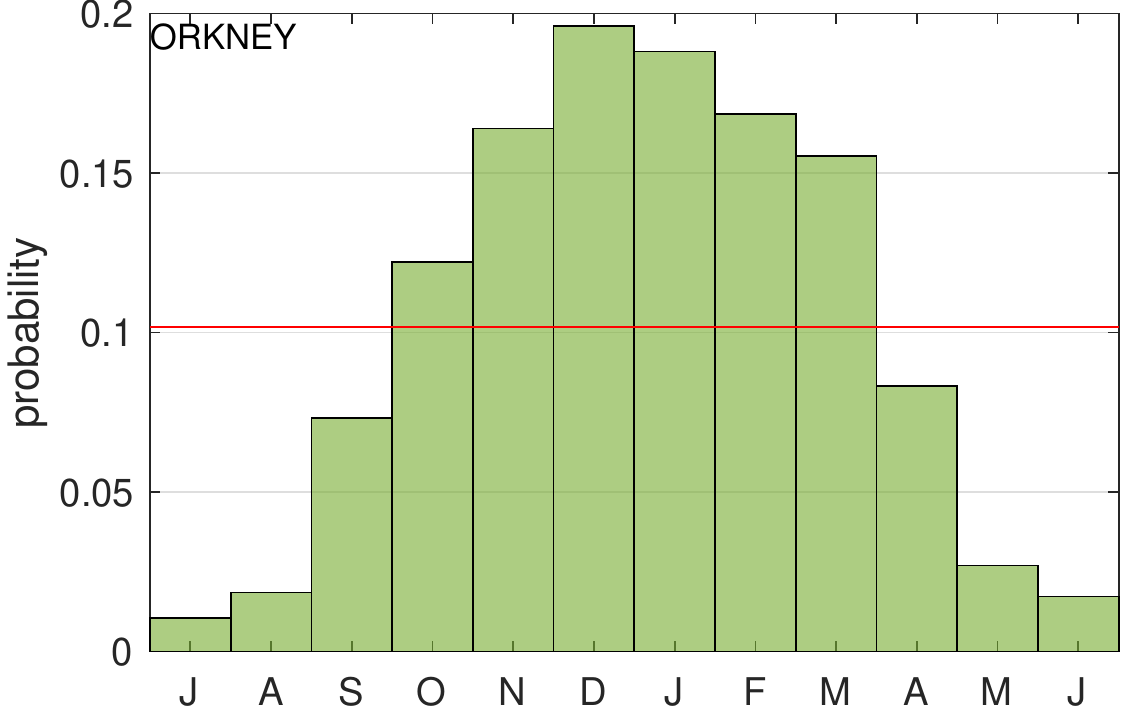}}
\caption{Bar chart showing proportion of each month during which the M4 power output is clipped, i.e. when the WEC operates at its maximum capacity. Albany plot is shown on the left and Orkney on the right. The horizontal lines represent the long-term average proportion. Note that for Albany the x-axis is from January to December, while for Orkney it is from July to June.}
\label{fig_histogram_off_clip}
\end{figure}

Our model assumes the WEC to be operational when the potential power yield would be above the machine rated power. Then, the power-take-off characteristics are assumed to be adjusted such that the maximum capacity is not exceeded. As such, power clipping corresponds to the machine operating at its maximum capacity. Figure \ref{fig_histogram_off_clip} shows a bar chart for the proportion of time, within each month, that the M4 wave energy converter is clipped throughout the duration of the hindcast data. Note that the x-axis for Orkney is shifted by 6 months, such that for Albany the x-axis is from January to December, while for Orkney it is from July to June. The horizontal line represents the mean annual probability of clipping. Accordingly, the power yield is clipped approximately 2.5\% of the time at Albany, while close to 10\% of the time at Orkney. In Albany clipping instances are extremely rare during summer, and in even in winter, power saturation occurs roughly only 6\% of time. On the other hand, the severe winter storms at Orkney result in approximately 18\% probability of power clipping during this season. Power clipping due to the finite machine capacity mitigates the extremes in the incident wave conditions and ultimately leads to lower seasonal variation in the production compared to the resource. This is illustrated in Figure \ref{fig_sizing} and briefly re-visited in Section \ref{sec_WEC_size}.


\subsection{Probability and persistence of low production}
\label{sec_low_P}
\begin{figure}[ht]
\centering
{\includegraphics*[width=0.49\linewidth, viewport=121 317 445 518]{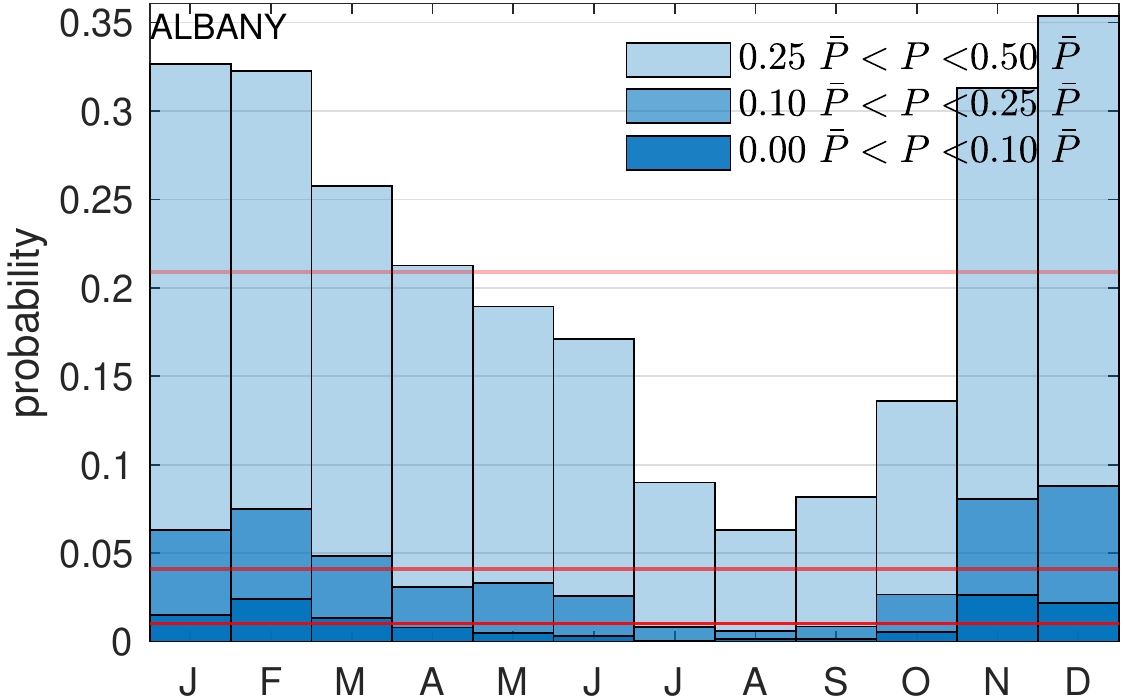}}
{\includegraphics*[width=0.49\linewidth, viewport=121 317 445 518]{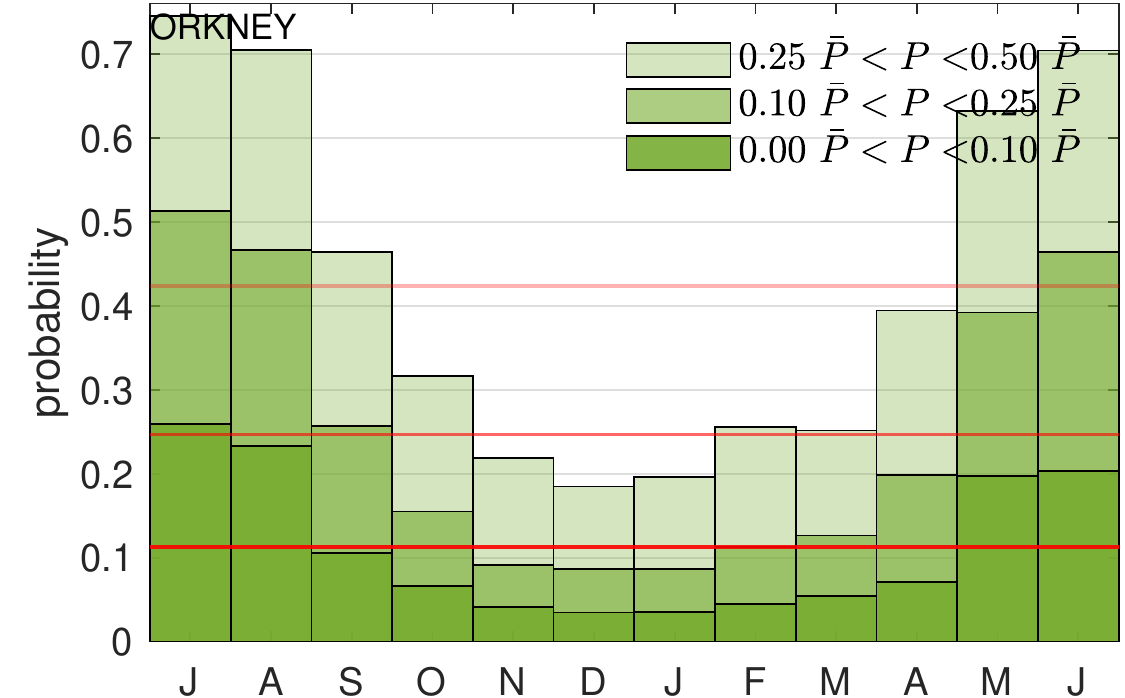}}
\caption{Bar chart showing proportion of each month during which the M4 power output is within threshold levels of 0.10, 0.25 and $0.50 \bar{P}$. Albany plot is shown on the left and Orkney on the right. The horizontal lines represent the long-term average probabilities of power output being below each power level threshold. Note that for Albany the x-axis is from January to December, while for Orkney it is from July to June.}
\label{fig_histogram_low_P}
\end{figure}

In terms of power supply reliability, it is of interest to investigate low-power occurrences. We choose three different power levels: 0.10, 0.25 and 0.50 $\bar{P}$ (approximately 3, 8 and 17\% of the machine capacity). Figure \ref{fig_histogram_low_P} shows a bar chart for the proportion of time, within each month, during which the power output is within these threshold levels. The horizontal lines represent the long-term average probabilities of the power output being below the threshold levels. At both locations, the intra-annual variation in low-power occurrences is evident. Overall, Orkney exhibits much higher probabilities of low-production than Albany. In Albany, the lowest chances of low-power events are in mid/late winter and early/mid spring (months of July - October, depending on the threshold power level). There is a considerable increase in low-production instances in November. This fast transition in mid/late spring is observable in the monthly wave resource and the monthly M4 yield plots (see the Albany plots in Figures \ref{fig_monthly_power}, \ref{fig_monthly_power_wave} and \ref{fig_sizing}) and presumably follows from large-scale geophysical phenomena affecting the local wave climate at the south-western Australian coast. On the other hand, in Orkney the intra-annual variation is close to symmetric about the highest chance of low-production in the months of May-August. In Albany the probability of occurrence of $P<0.10 \bar{P}$ is very low; just over 1\% annually, while it reaches 11\% at Orkney, pointing to a much more intermittent power supply. Power levels below $0.50 \bar{P}$ occur roughly 21\% of the time annually in Albany, and are more common in the summer (with the highest probability in the month of December, corresponding to roughly 11 days out of 31). In Orkney such low-production levels occur approximately 24 out of 31 days in July, and on average 42\% of the time.
\begin{figure}[ht]
\centering
{\includegraphics*[width=0.99\linewidth, viewport=60 339 528 501]{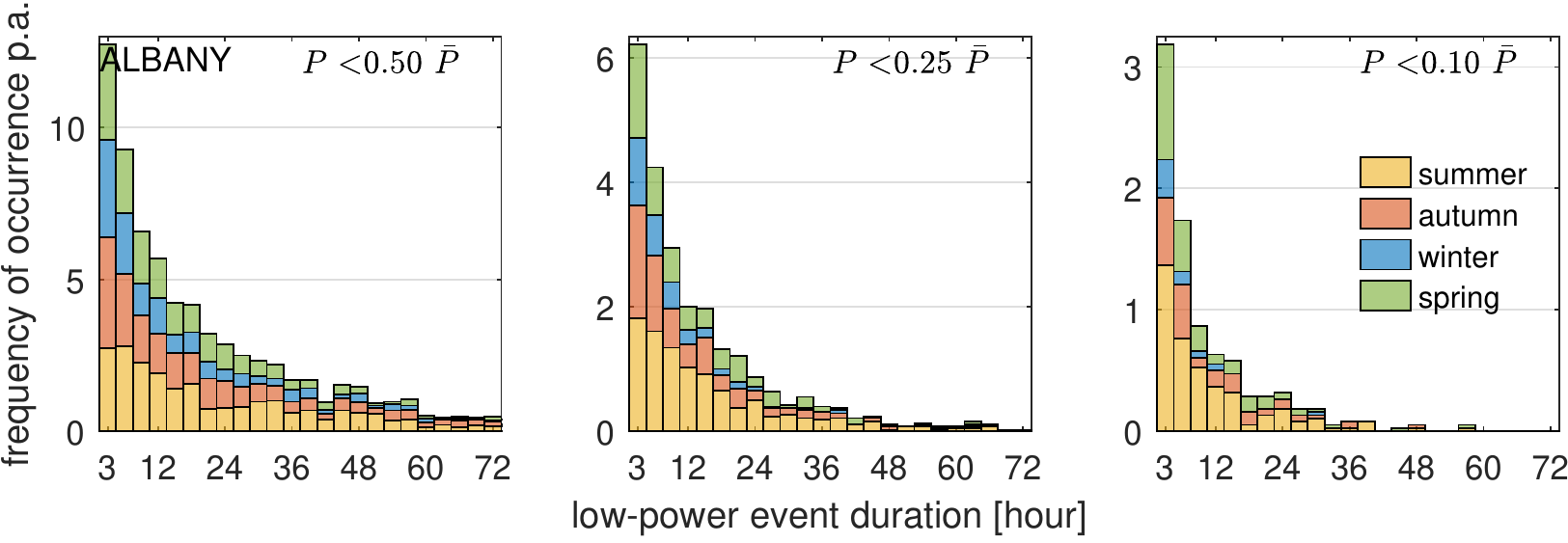}}
{\includegraphics*[width=0.99\linewidth, viewport=60 339 528 501]{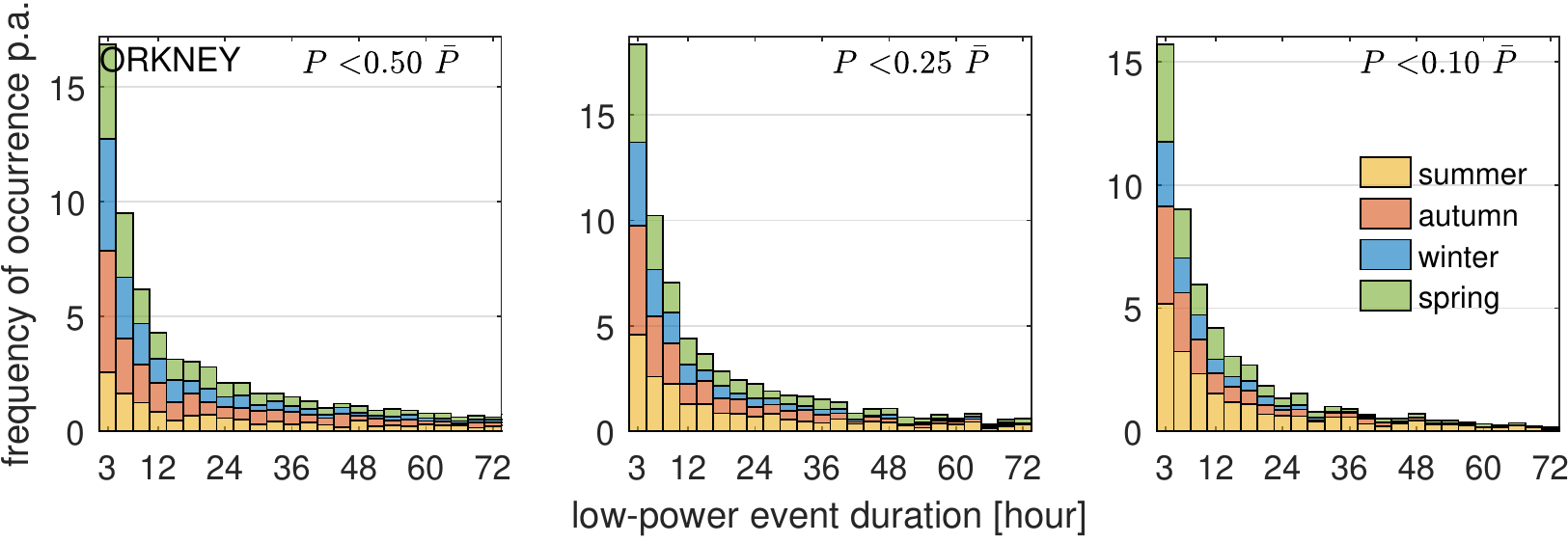}}
\caption{Frequency and duration of low-power events, for each season, assuming threshold values of $0.10$, $0.25$ and $0.50\bar{P}$. The bars are stacked, such that the total bar height represents annual frequency of occurrence. Albany plot is shown at the top and Orkney at the bottom.}
\label{fig_low_P_dur_freq_seasons}
\end{figure}

In addition to the probability of occurrence of low production, we also quantify its persistence. We define low-power persistence as the duration of consecutive 3-hourly intervals with power output below the three chosen threshold levels of $0.10$, $0.25$ and $0.50 \bar{P}$. Figure \ref{fig_low_P_dur_freq_seasons} presents different duration low-power events and their respective frequency of occurrence in a year. The overall height of each bar represents the number of the low-power events per year, and how these events are partitioned into the four seasons is indicated by the different colours (the different seasons are effectively stacked on top of each other). Note that the graphs are truncated at 3-days long events; longer duration events are discussed separately (see Table \ref{tab_return_period_low_P}). Overall, shorter duration low-power events are more common, while longer events become progressively more rare. The displayed distributions of $P<0.50 \bar{P}$ events appear similar at the two locations; for example, annually there are approximately 34 and 36 events of duration up to 12 hours, 14 and 11 events of duration between 1 and 2 days, and 5 and 6 events of duration between 2 and 3 days in Albany and in Orkney respectively. However, the tails of the distributions deviate (not shown); for example, events over 5 days long occur on average 7 times annually in Orkney while just over once per year in Albany. Lower power events are much more persistent at Orkney compared to Albany; for example, annually there are 35 events up to 12-hours long with $P<0.10 \bar{P}$ at Orkney, while only 6 such events at Albany. From Figure \ref{fig_low_P_dur_freq_seasons} it follows that seasonal effects are not strikingly discernible for $P<0.50 \bar{P}$, while low-power events with $P<0.10 \bar{P}$ are visibly more common in summer compared to winter. We note in passing that similar persistence analyses have been carried out by \cite{holttinen2005}, \cite{Cannon2015extreme_wind} and \cite{Ohlendon2020_freq_dur_low_wind} to assess low wind power production in various European countries.


Table \ref{tab_return_period_low_P} summarises a number of extracted values from the persistence analysis for easier comparison of the two locations. Instead of the annual frequency of occurrence as in Figure \ref{fig_low_P_dur_freq_seasons}, we use a return period, expressed in years, which we simply define as the reciprocal of the number of events per year. The last row of the table contains the longest recorded low-production events in the two hindcast datasets. At any given return level, low-power events are significantly more persistent at Orkney compared to Albany (apart from once-per-year $P<0.50 \bar{P}$ events). The lowest production $P<0.10 \bar{P}$ periods are 3 - 5 times shorter at Albany, indicating a more continuous power supply.
\begin{table}[ht!]
\begin{center}
\begin{tabular}{|c || c c c|| c c c||} 
\hline
& \multicolumn{6}{|c||}{Duration in hours (days)} \\
\hline
& \multicolumn{3}{|c||}{Albany} &\multicolumn{3}{|c||}{Orkney}\\
\hline
Return period in years & $P<0.50 \bar{P}$ & $P<0.25 \bar{P}$ & $P<0.10 \bar{P}$ & $P<0.50 \bar{P}$ & $P<0.25 \bar{P}$ & $P<0.10 \bar{P}$ \\ 
\hline
1           &  57 (2.4)  &  21 (0.9) &  6 (0.3) &  48 (2.0)  &   48 (2.0)  &   33 (1.4)\\ 
\hline
2           &  78 (3.3)  &  33 (1.4) & 15 (0.6) &  96 (4.0)  &   72 (3.0)  &   48 (2.0)\\
\hline
5           & 105 (4.4)  &  45 (1.9) & 24 (1.0) & 189 (7.9)  &  144 (6.0)  &   78 (3.3)\\
\hline
10          & 138 (5.8)  &  66 (2.8) & 30 (1.3) & 279 (11.6) &  180 (7.5)  &  126 (5.3)\\
\hline
38 or 54    & 291 (12.1) & 117 (4.9) & 93 (3.9) & 297 (12.4) &  297 (12.4) &  231 (9.6)\\ 
\hline
\end{tabular}
\end{center}
\caption{Duration of low-power events in hours (values in brackets represent approximate duration in days) for different return periods in years, assuming power threshold values of $0.10$, $0.25$ and $0.50\bar{P}$. The bottom row represents the longest recorded low-power events in the Albany 38-year hindcast (1980--2017) and in the Orkney 54-year hindcast (1958--2011).}
\label{tab_return_period_low_P}
\end{table}

\subsection{Monthly distribution}
\label{sec_months}
\begin{figure}[ht!]
\centering
{\includegraphics*[width=0.49\linewidth, viewport=120 315 476 524]{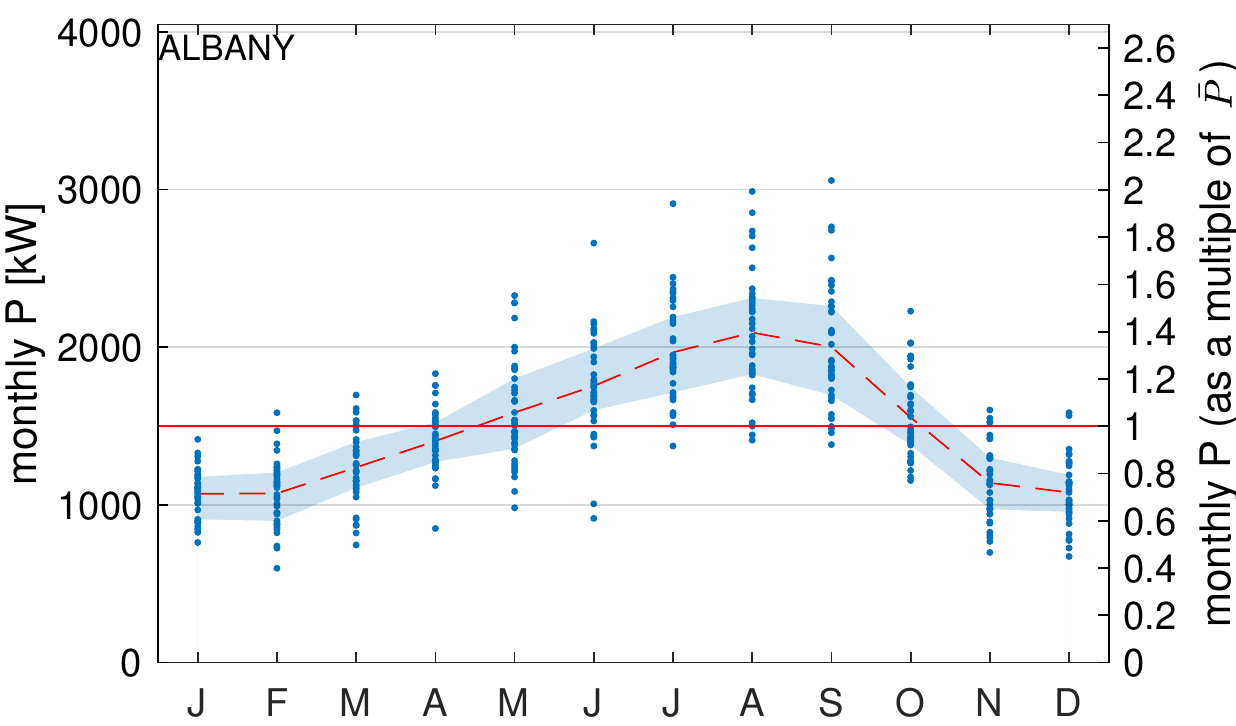}}
{\includegraphics*[width=0.49\linewidth, viewport=120 315 476 524]{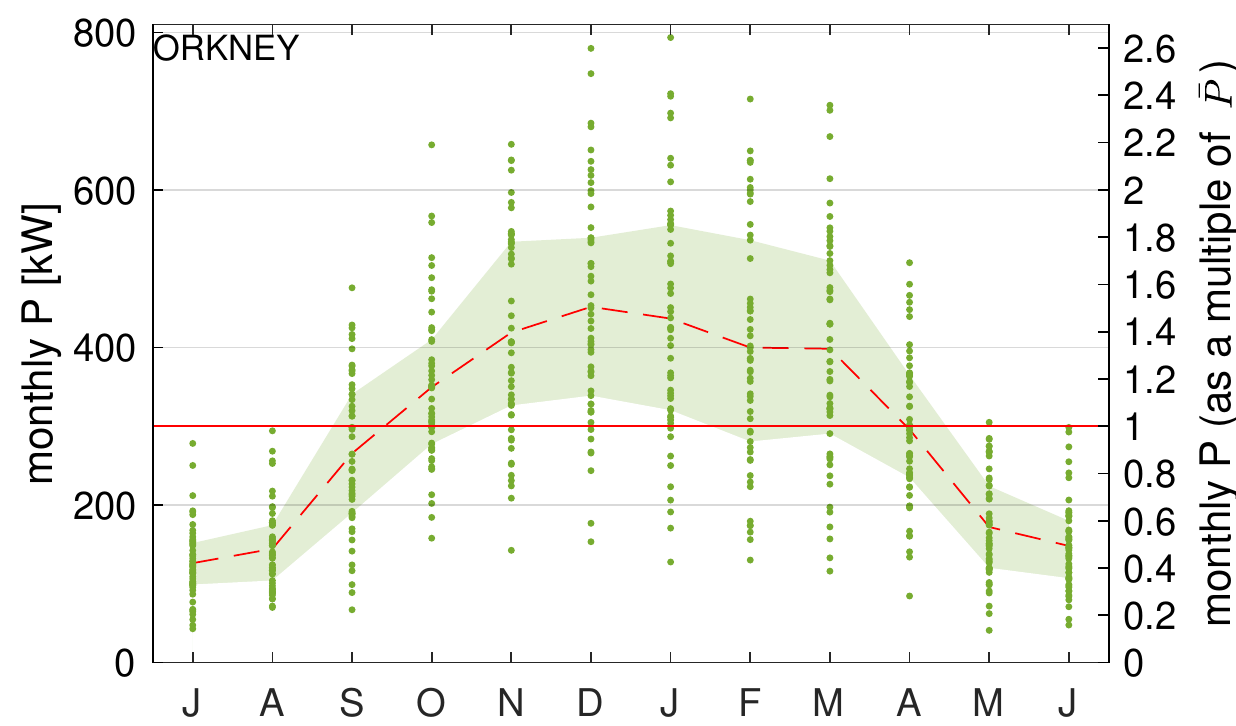}}
\caption{Distribution of individual monthly mean absorbed power (dot markers). Albany plot is shown on the left and Orkney on the right. The long-term average absorbed power $\bar{P}$ is shown by the solid red line, while the dashed red line represents the long-term average for each calendar month. The shaded area demarcates the first and third quartiles of each monthly distribution (i.e. it represents the middle 50\% of the data). Note that for Albany the x-axis is from January to December, while for Orkney it is from July to June.}
\label{fig_monthly_power}
\end{figure}

Having analysed the 3-hourly distribution of practical wave power, we now present the results in terms of monthly distribution. We calculate the mean absorbed power for every calendar month in the hindcast datasets, and plot the distribution of these monthly mean values in Figure \ref{fig_monthly_power} using dot markers. We adopt the same plotting convention as in Figure \ref{fig_histogram_monthly}. The red dashed line represents the long-term average value for each month (i.e. the average of the individual monthly means) and as such shows the intra-annual variation of the monthly power and corresponds to the vertical dashed lines in Figure \ref{fig_histogram_monthly}. The red solid line represents the long-term mean absorbed power $\bar{P}$. A seasonal cycle is observed at both locations. The right-hand-side y-axis displaying the relative power levels allows for easy comparison. Note that the axis limits are deliberately chosen to be the same for both locations. At Albany the intra-annual variation of the long-term average monthly power values is within 0.7 and 1.4 $\bar{P}$, while at Orkney the winter levels are 3 times higher than in the summer. 

\begin{figure}[ht!]
\centering
{\includegraphics*[width=0.49\linewidth, viewport=127 315 444 522]{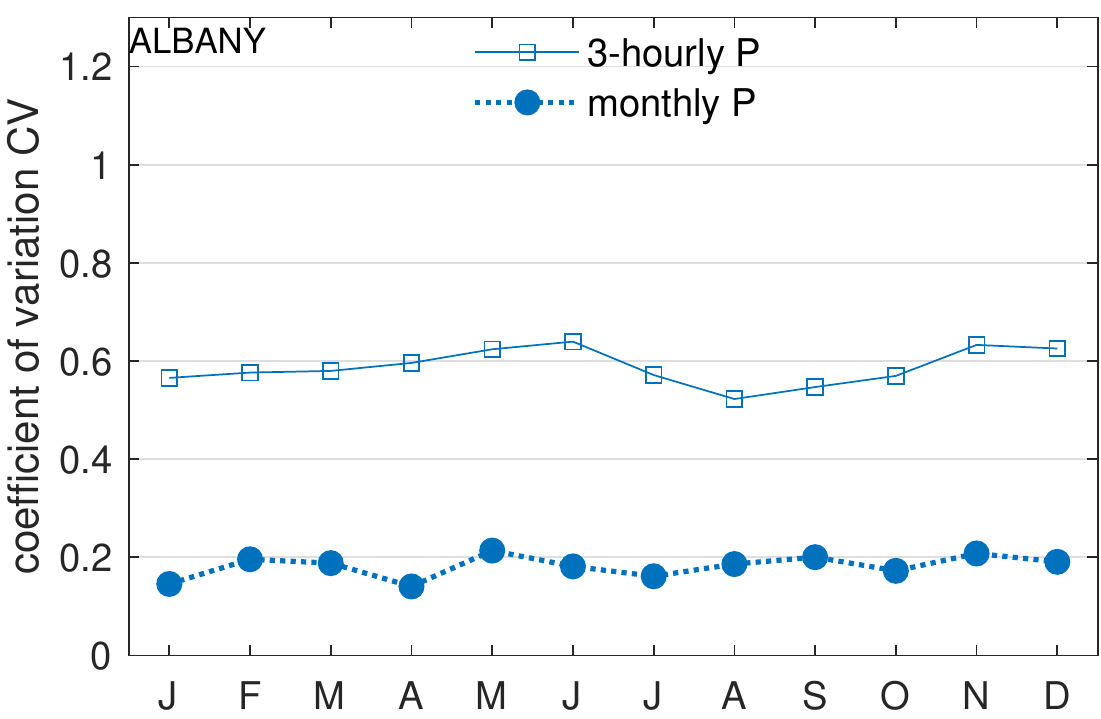}}
{\includegraphics*[width=0.49\linewidth, viewport=127 315 444 522]{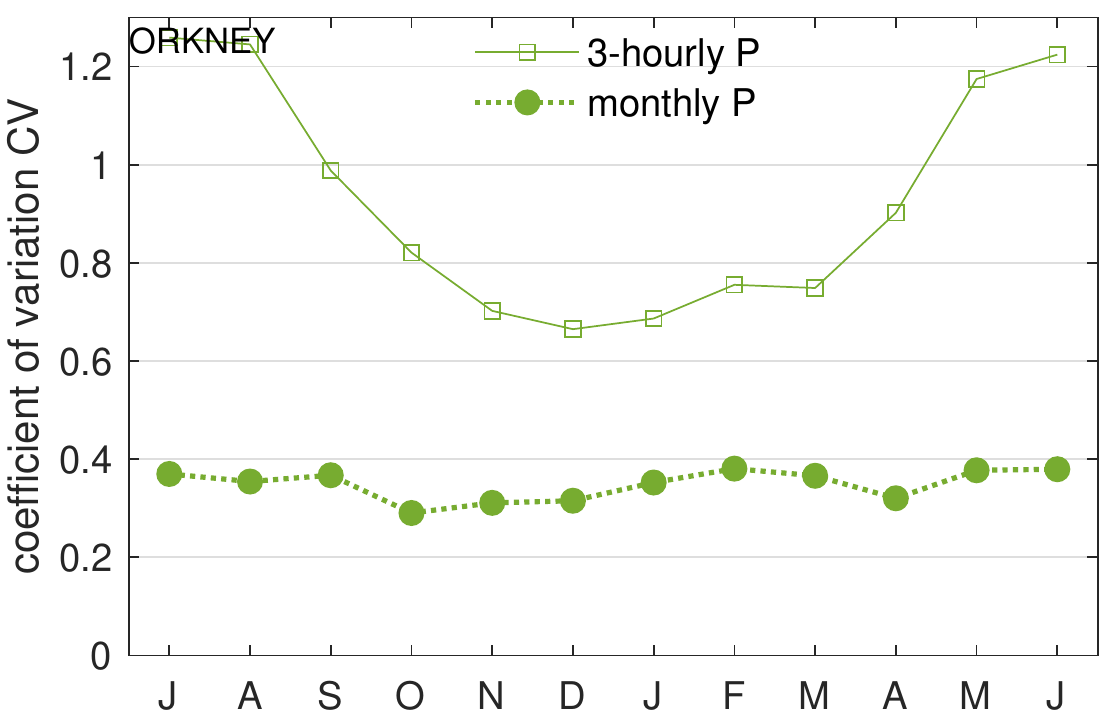}}
\caption{Coefficient of variation for 3-hourly and monthly absorbed power values. Albany plot is shown on the left and Orkney on the right. Note that for Albany the x-axis is from January to December, while for Orkney it is from July to June.}
\label{fig_CV}
\end{figure}

Despite the large seasonal variation, it is interesting to note that the variability of the individual monthly mean wave power is close to symmetric about the average value. This is illustrated by the shaded area on the plot which demarcates the first and third quartiles of the monthly distributions (i.e. it represents the middle 50\% of the data). To measure the degree of variability for each month, the coefficient of variation (CV) is used, defined as the ratio of the standard deviation to the mean. This is shown in Figure~\ref{fig_CV} for the monthly distributions, as well as for the 3-hourly distributions plotted in Figure~\ref{fig_histogram_monthly}. It is apparent that when the practical wave power is assessed in terms of monthly values, at both locations the variability is quite uniform throughout the twelve months without any obvious seasonal effects. This has potential implications for power forecasting, such that if one was able to predict the monthly mean absorbed power with some confidence in a particular month, January say, then one would expect the same confidence level to be applicable to the rest of the months. However, when assessed in terms of 3-hourly power values, there is a strong seasonal cycle at Orkney with over 70\% more variability in summer than in winter. In Albany, there is no such discernible seasonal trend. Overall, whether in terms of the shorter-term 3-hourly or the longer-term monthly production, the variability is considerably lower in Albany, suggesting that the power output is more reliable (and potentially more predictable). 
\begin{figure}[ht!]
\centering
{\includegraphics*[width=0.49\linewidth, viewport=120 315 476 529]{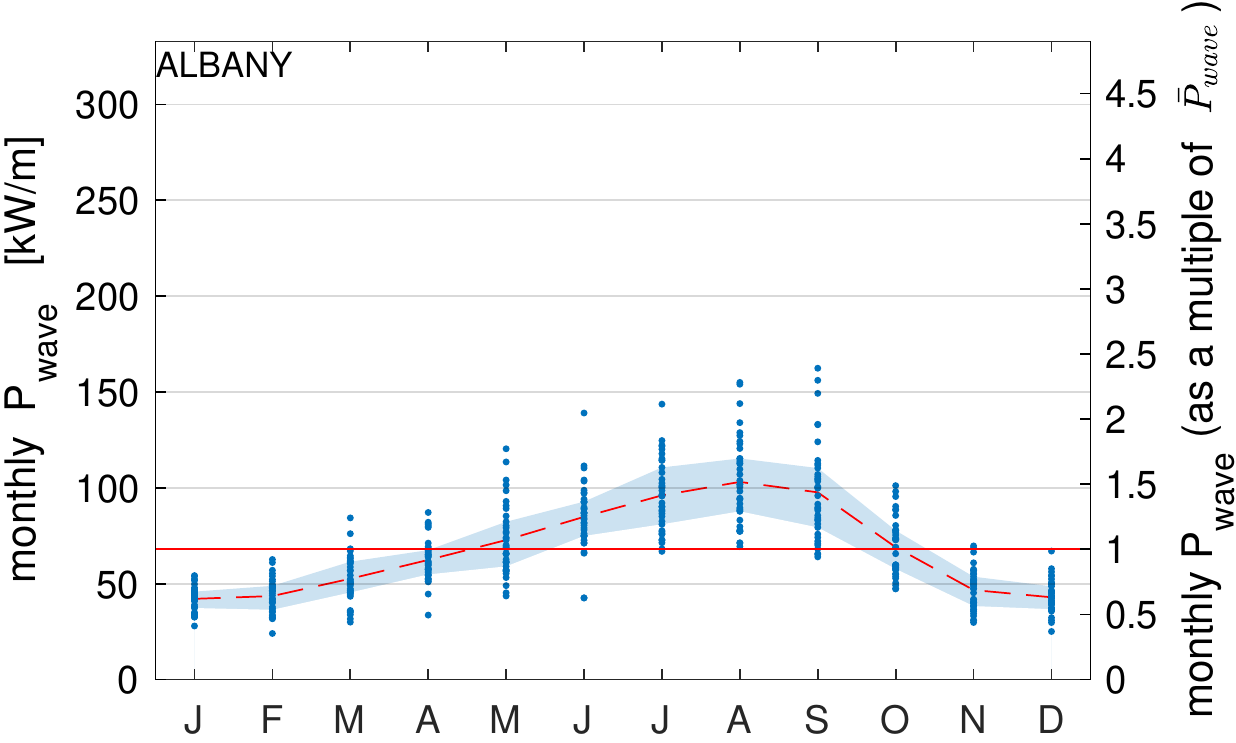}}
{\includegraphics*[width=0.49\linewidth, viewport=120 315 476 529]{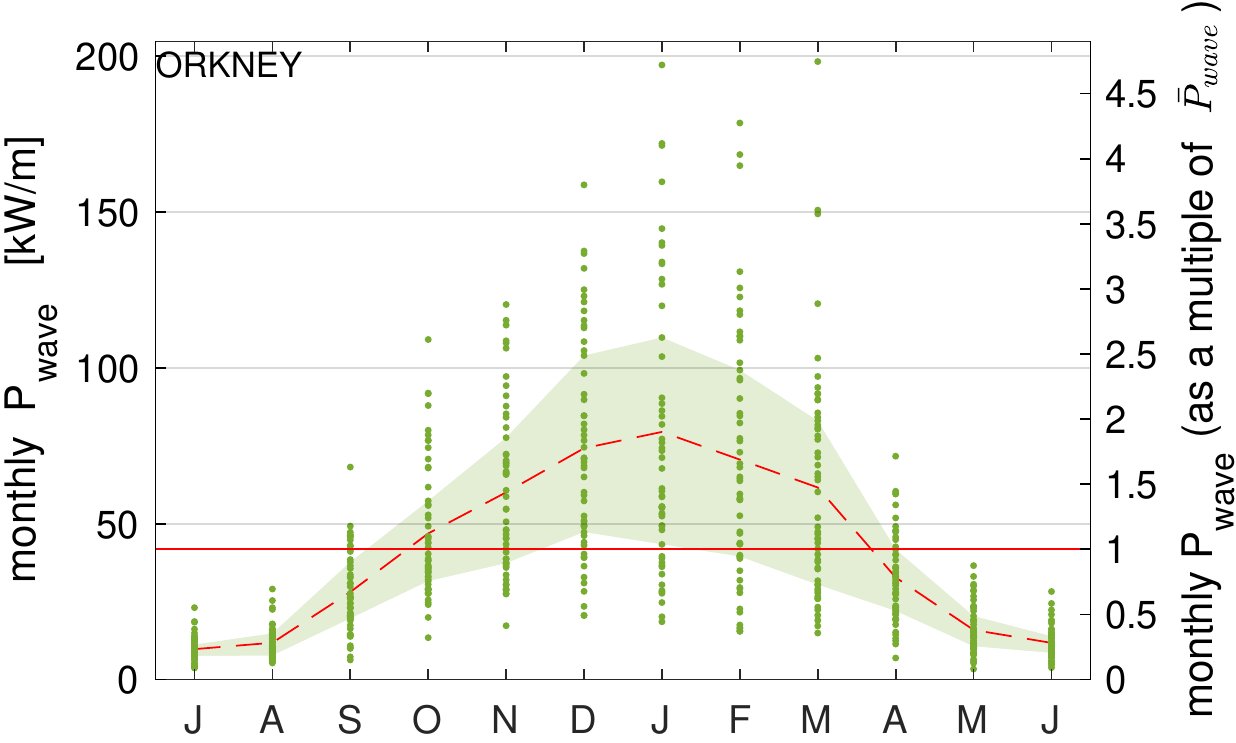}}
\caption{Distribution of individual monthly mean incident wave power density (dot markers). Albany plot is shown on the left and Orkney on the right. The long-term average wave power density $\bar{P}_{wave}$ is shown by the solid red line, while the dashed red line represents the long-term average for each calendar month. The shaded area demarcates the first and third quartiles of each monthly distribution (i.e. it represents the middle 50\% of the data). Note that for Albany the x-axis is from January to December, while for Orkney it is from July to June.}
\label{fig_monthly_power_wave}
\end{figure}

For reference, in Figure \ref{fig_monthly_power_wave}, we show the distribution of monthly mean values of the incident wave power density. We note the larger seasonal variation of the long-term average monthly wave resource values compared to the absorbed power shown in Figure \ref{fig_monthly_power}. This is much more pronounced at Orkney, where the average intra-annual resource variations are within 0.2 - 1.9 of the long-term mean wave energy flux density, such that there is close to 7 times more incident wave energy in winter compared to summer. As discussed before, the machine capture width characteristic and power clipping reduce the degree of temporal variability in the absorbed power. Similarly, the inter-annual (i.e. year-to-year) variability is larger for the resource than for the power production. For Albany the CV of the annual incident wave energy flux densities is roughly 0.11, while the CV for the annual absorbed power is 0.09. For Orkney the corresponding CVs are 0.22 and 0.13 respectively. It is also worth noting that, because of the effect of aggregation, the fluctuations in the annual values are smaller than the monthly variations, which are yet smaller than the 3-hourly variations.

\section{Effect of machine sizing}
\label{sec_WEC_size}
\begin{figure}[ht!]
\centering
{\includegraphics*[width=0.49\linewidth, viewport=120 315 444 522]{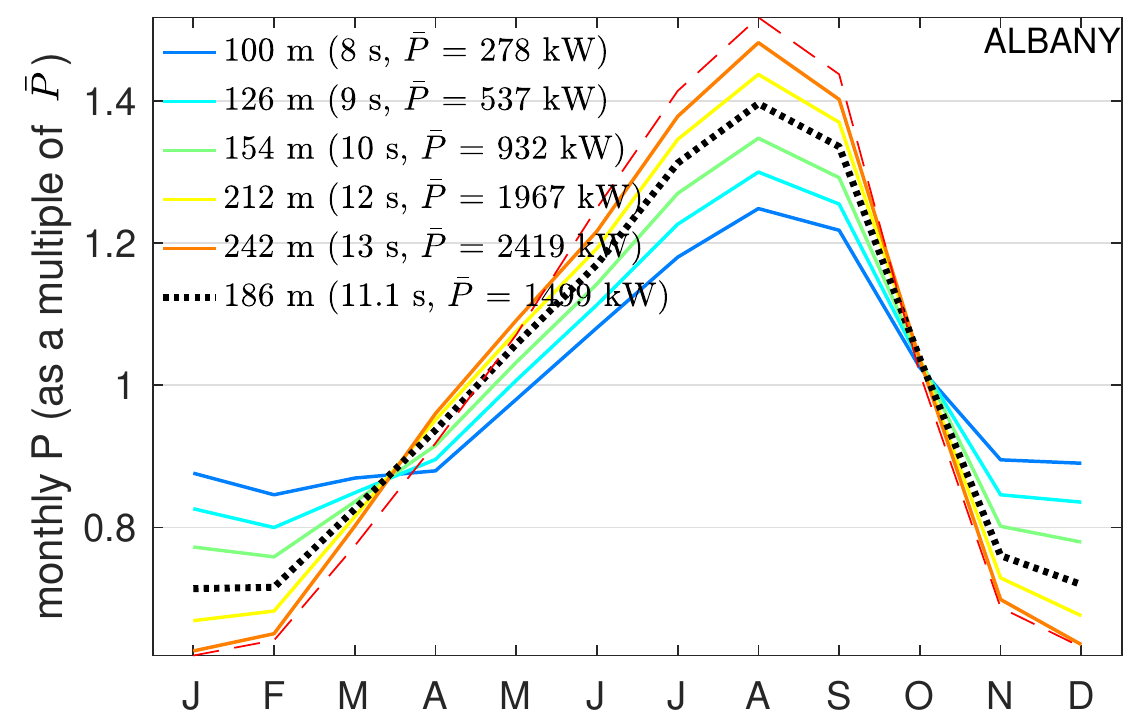}}
{\includegraphics*[width=0.49\linewidth, viewport=120 315 444 522]{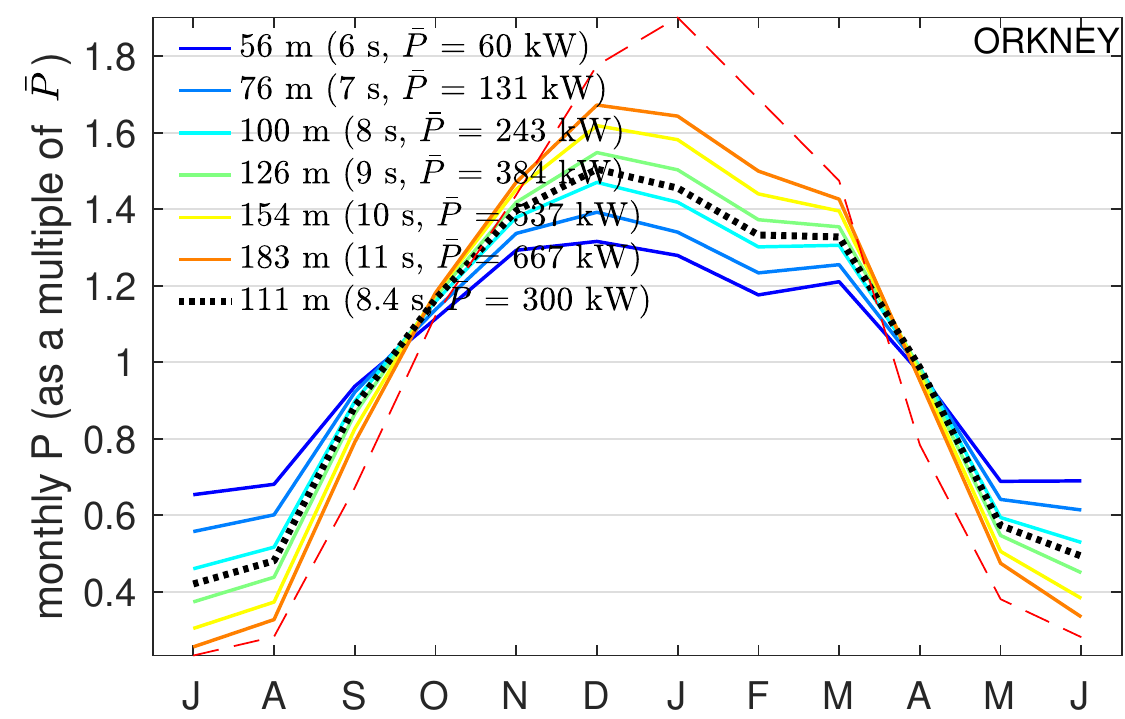}}
\caption{Long-term average monthly absorbed power for different size M4 machines. Albany plot is shown on the left and Orkney on the right. The monthly power values have been normalised by the corresponding long-term averages $\bar{P}$. The WEC sizes are shown in the legend (text in the brackets represents the corresponding wave period and $\bar{P}$). The normalised long-term average monthly wave resource, as per Figure \ref{fig_monthly_power_wave}, is shown by the dashed red line. Note that for Albany the x-axis is from January to December, while for Orkney it is from July to June.}
\label{fig_sizing}
\end{figure}

It is of interest to investigate the effect of machine size on the variability of the M4 production performance. We consider a range of machine lengths corresponding to wavelengths of $8-13$ s waves for Albany, and of $6 - 11$ s waves for Orkney. We recall that the standard size machines are tuned to the long-term average energy period, such that the WEC length (from the middle of the bow float to the middle of the stern float) is set according to the corresponding wavelength. For Albany and Orkney, the long-term $T_e$ values are 11.1 and 8.44 s respectively, with the standard lengths being 186 and 111 m. As shown in Figure \ref{fig_sizing}, machine size considerably affects the seasonal variation of the absorbed power. For both locations, as the machine becomes longer, the variation increases, and the intra-annual trend approaches that of the wave resource (dashed red line, taken from Figure \ref{fig_monthly_power_wave}). Smaller machines exhibit less seasonal fluctuations. For the smallest machines considered, the lowest monthly means are approximately 0.85 and 0.65$\bar{P}$ for Albany and Orkney respectively. However, unsurprisingly, the mean power yield, in terms of kW, diminishes significantly as the machine size is reduced.
\begin{figure}[ht!]
\centering
{\includegraphics*[width=0.99\linewidth, viewport=29 339 574 513]{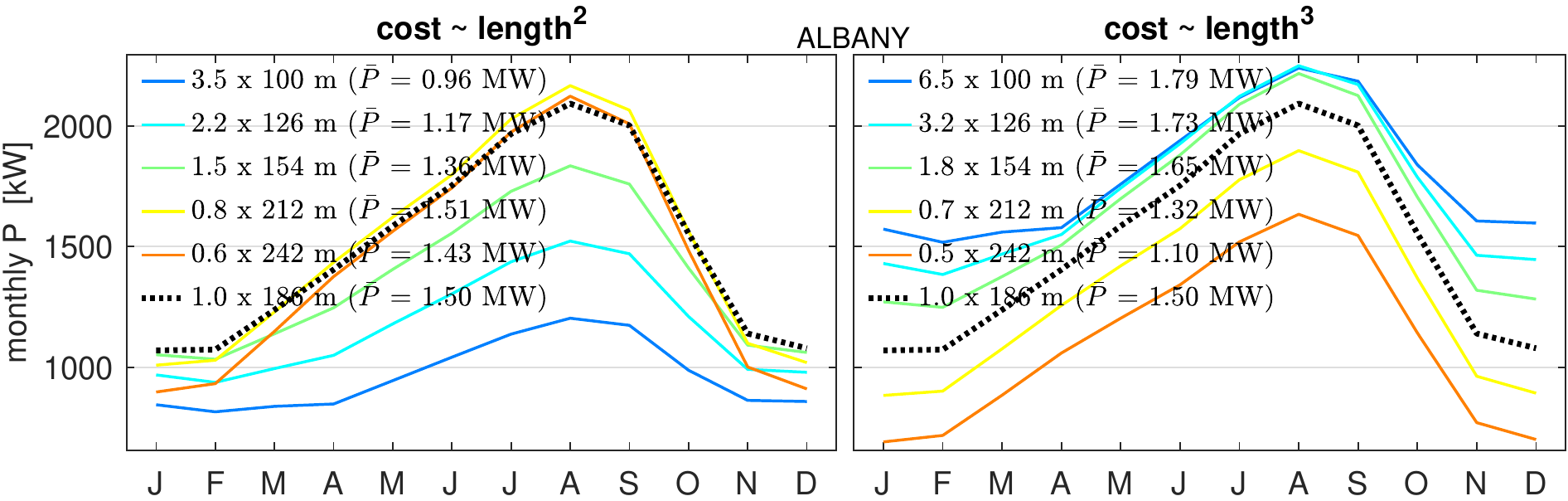}}
{\includegraphics*[width=0.99\linewidth, viewport=29 339 574 513]{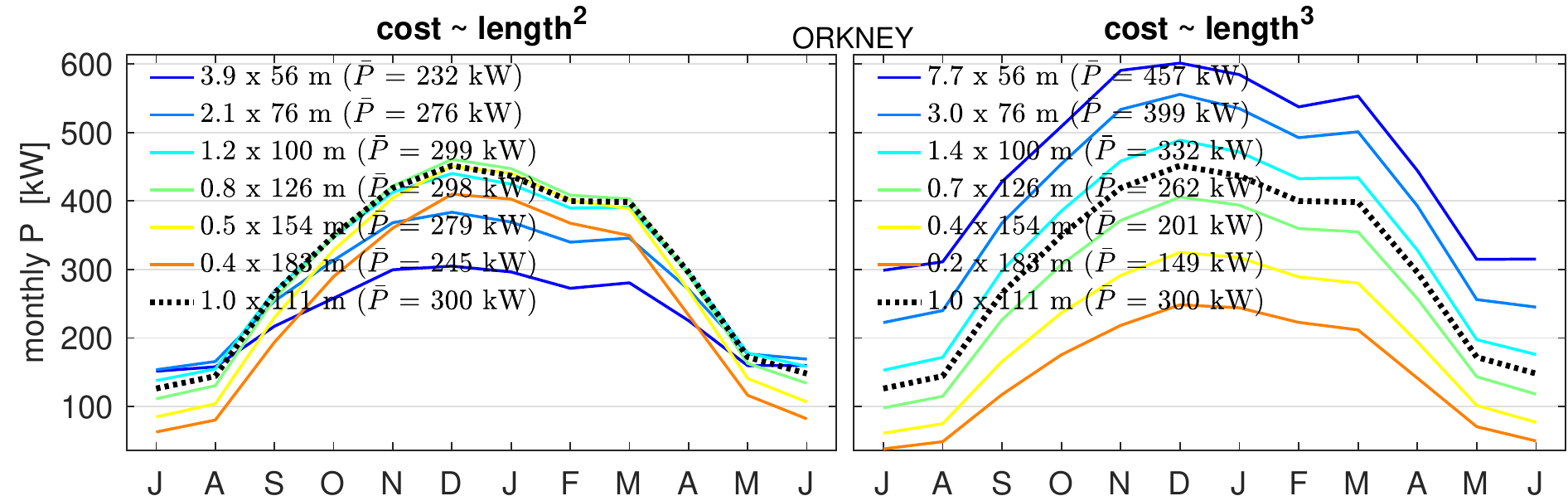}}
\caption{Long-term average monthly absorbed power for different size M4 machines, with the number of machines at each size selected to give the same total cost as a single standard machine. Albany plot is shown at the top and Orkney at the bottom. The legend shows the number of machines times the machine length (text in the brackets represents the total long-term average power output). Note that for Albany the x-axis is from January to December, while for Orkney it is from July to June.}
\label{fig_sizing_cost}
\end{figure}


We now carry out simplified cost calculations to compare different size M4 WECs in terms of economics of their production. Following \cite{santo2016decadalM4} and \cite{santo2020_M4_EMEC_Albany} we assume the total production costs to be proportional to the machine length raised to power 2 and 3. Such a simple cost function follows from assumptions that some costs will likely be proportional to the surface area and the mass, thereby giving rise to the square and the cube dependence. In reality, of course, the economic analysis is much more involved, with many cost components (especially installation and operational costs) not obeying a simple relationship to the WEC size. Figure \ref{fig_sizing_cost} shows the monthly power yields from multiple units of different size machines. For each size, the number of machines is chosen such that the total cost is equivalent to the cost of the standard M4 unit (at that location); this is indicated in the legend of the plot. For both locations, assuming length$^2$ cost dependence, the standard machines appear optimal. Even though deploying a slightly bigger device (see for example the 212 m long machine in Albany and the 126 m long machine at Orkney) would result in comparable production performance (in terms of the economics of the overall power yield $\bar{P}$), the output would be slightly more seasonal and more adversely affected by intermittency (see Table \ref{tab_return_period_low_P_diff_size}). Moreover, assuming the cubic cost function, which is perhaps a more realistic economic model, especially for early deployments whose costs will be high, larger devices are found to be less cost-effective. Overall, larger devices do not appear to be favourable. At both locations, reducing the machine size is seen to be beneficial, under the cubic cost model, as smaller devices deliver higher $\bar{P}$ for the same cost, are less prone to seasonal effects and exhibit shorter and less frequent low-power events (see Table \ref{tab_return_period_low_P_diff_size}). Our overall recommendation of smaller devices is supported by \cite{coe2021} for example, who suggest such devices might be preferred due to higher yields and higher capacity factors. We note that the cost comparison of different device sizes presented above in Figure \ref{fig_sizing_cost} is equivalent to normalising (i.e. dividing) the power output of each single device by the square or the cube of its length. We do not consider device performance expressed as the power yield per unit length (i.e. cost $\sim$ length) because machine length is not a realistic indicator of cost.


To complement the above discussion, Table \ref{tab_return_period_low_P_diff_size} summarises low-production persistence and frequency of occurrence for $P<0.10 \bar{P}$ events. For each location, the results for the standard size machine (as per Table \ref{tab_return_period_low_P}) are shown together with approximately 68\% smaller and 114\% bigger devices. For Albany, the machines lengths are 126, 186 and 212 m, corresponding to 9, 11.1 and 12 s waves. For Orkney, the machine lengths are 76, 111 and 126 m corresponding to 7, 8.4 and 9 s waves. For any return level, low-power events are considerably shorter for smaller devices. This is true for Albany in particular, where power intermittency is drastically reduced with a smaller device. 
\begin{table}[ht!]
\begin{center}
\begin{tabular}{|c || c c c|| c c c||} 
\hline
& \multicolumn{6}{|c||}{$P<0.10 \bar{P}$ duration in hours (days)} \\
\hline
& \multicolumn{3}{|c||}{Albany} &\multicolumn{3}{|c||}{Orkney}\\
\hline
Return period in years & $L_{M4}$=126 m  & $L_{M4}$=186 m & $L_{M4}$=212 m & $L_{M4}$=76 m & $L_{M4}$=111 m & $L_{M4}$=126 m \\ 
\hline
1           &  0 (0.0)   &  6 (0.3) &  12 (0.5) &  21 (0.9) &  33 (1.4) &  30 (1.3)\\ 
\hline
2           &  0 (0.0)   & 15 (0.6) &  18 (0.8) &  30 (1.3) &  48 (2.0) &  63 (2.6)\\
\hline
5           &  6 (0.3)   & 24 (1.0) &  36 (1.5) &  57 (2.4) &  78 (3.3) & 105 (4.4)\\
\hline
10          &  6 (0.3)   & 30 (1.3) &  48 (2.0) &  75 (3.1) & 126 (5.3) & 108 (4.5)\\
\hline
38 or 54    & 27 (1.1)   & 93 (3.9) &  93 (3.9) & 225 (9.4) & 231 (9.6) & 267 (11.1)\\ 
\hline
\end{tabular}
\end{center}
\caption{Duration of $P<0.10\bar{P}$ low-power events in hours (values in brackets represent approximate duration in days) for different return periods in years, for three different M4 WEC sizes (approximately 0.68, 1.00 and 1.14 times the standard machine length at each location). The bottom row represents the longest recorded low-power events in the Albany 38-year hindcast (1980--2017) and in the Orkney 54-year hindcast (1958--2011).}
\label{tab_return_period_low_P_diff_size}
\end{table}

\section{Discussion and Conclusions}
In this work we analysed temporal variability of predicted power production of an M4 wave energy converter (WEC) deployed at two energetic offshore locations; off Albany on the south-western coast of Australia and off the European Marine Energy Centre (EMEC) at Orkney, UK. At both locations long-term hindcast wave data was utilised, each spanning approximately 2-3 hypothetical lifespans of an offshore energy installation. The WEC size was chosen according to the long-term average energy period of each wave climate. A frequency-domain hydrodynamic model was then applied to reconstruct historical M4 power production, followed by statistical analysis considering 3-hourly, monthly, seasonal and inter-annual timescales.

We first investigated probability distributions of 3-hourly power $P$ and found these to be positively skewed. For Orkney this asymmetry was much more pronounced indicating low production to be more likely (for example in Orkney almost a quarter of the time the production was $P \le 0.25 \bar{P}$, with $\bar{P}$ denoting the long-term average power yield, while such conditions occurred on average only 4\% of the time in Albany). Albany exhibited a narrower distribution, corresponding to less variable production compared to Orkney. This was also reflected in the coefficient of variation (a non-dimensional measure of spread in the data) for the 3-hourly power, which was 0.6 for Albany and 0.9 for Orkney. 

To elucidate intermittency of the production, we analysed low-power instances considering three threshold levels of $P< (0.50, \ 0.25, \ 0.10) \bar{P}$. As expected, low-production was more likely in summer (most common in November-February in Albany and in May-August in Orkney). We quantified persistence of low-power events and their annual frequency of occurrence. Considering the minimal threshold investigated, $P<0.1\bar{P}$, for all return levels considered, low-production events were found to be 3-5 times longer at Orkney (for example a typical once-per-year $P<0.1\bar{P}$ event was found to be 6-hours long at Albany, while it would be 33-hours long at Orkney). Comparing the two locations, Orkney exhibited much more intermittent production with longer and more frequent occurrences of minimal power output.

To quantify seasonal variations we calculated monthly absorbed power yields. The averaged intra-annual trends from both locations suggested higher seasonal fluctuation at Orkney. This was found to stem from the resource; in Orkney there was almost 7 times more incident wave energy in winter compared to summer, whereas in Albany this ratio was slightly over 2. The winter extremes and the summer lows of the wave climate at Orkney were partially compensated by the finite operational (frequency) bandwidth of the WEC and by power clipping (arising from a finite physical capacity of the machinery). However, overall, the seasonality of the power output was still much more pronounced at Orkney compared to Albany, which benefits from a consistent year-round swell.

In summary, Albany was found to produce less intermittent and more steady output on 3-hourly, seasonal and inter-annual timescales. Even though not analysed in the present work, we additionally remark on survivability of a potential WEC deployment at the two locations considered. As per \cite{Barstow2009} and \cite{requero2015}, the ratio of the 100-year return period significant wave height and its long-term mean can be used as a metric for WEC deployment risk arising from extreme wave conditions. The calculated values are 3.3 and 5.2 for Albany and Orkney respectively (taken from \cite{santo2020_M4_EMEC_Albany}) pointing to another favourable characteristic of the Albany site. We note that metrics encompassing production performance, survivability and variability (amongst others) have been recently proposed by \cite{Lavidas2020_SIWED} and \cite{Kamranzad2020} for comprehensive assessment of WEC deployments. 



Lastly, machine sizing was also investigated and shown to significantly affect production levels and variability. At both locations, larger (than standard) devices generated more power on average, but were considered less cost-effective. Their power yield was found to be more intermittent and more seasonal - in fact the intra-annual variation approached that of the resource as the machine size was increased. Overall, smaller devices were found to be more favourable as they exhibited much reduced seasonal variations and intermittency. Our recommendation for smaller devices is not new, and has been advocated by wave energy experts for decades; for example, \cite{Falnes1993} proposed a WEC design principle of less variation in the power output than in the natural wave energy transport, which he suggested could be realised with smaller devices producing higher quality output. Similarly, \cite{Farley2012} highlighted the advantages of targeting shorter waves, which are more constant and steeper. We conclude that designing a WEC based on the average $T_e$ is not necessarily optimal, and the overall economics of different size devices in view of the variability of the wave climate needs to be considered. As an extension of the present work, it would be interesting to compare the M4 performance at the two studied energetic locations to areas with more benign but less variable wave climate, as a number of wave resource and WEC feasibility studies suggest these to be favourable for wave energy exploitation (see for example \cite{Portilla2013} and \cite{Fairley2020}).

\section*{CRediT authorship contribution statement} Writing - Original Draft: JO, HS and AK. Writing - Review \& Editing: all authors. Supervision: PHT, JO, AK and WZ. Conceptualisation: PHT and HS. Formal analysis: SL, JO and HS. Software, Resources and Data: MC, NM. SL was a visiting student at UWA. MC provided the Albany hindcast wave data set. NM and AK provided the M4 performance curves.   

\section*{Declaration of competing interest}
The authors declare that they have no known competing interests.

\section*{Acknowledgements}
Parts of this research were conducted by the Marine Energy Research Australia (MERA) and jointly funded by The University of Western Australia and the Western Australian Government, via the Department of Primary Industries and Regional Development (DPIRD). The support of the Blue Economy Cooperative Research Centre through the project `Seeding Marine Innovation in WA with a Wave Energy Deployment in Albany' is gratefully acknowledged. We thank Dr. Richard Gibson, then at BP Sunbury and now at Offshore Consulting Group, for providing the Orkney wave data. The Albany hindcast simulations were conducted on resources supported by the Pawsey Supercomputing Centre with funding from the Australian Government and the Government of Western Australia. HS was funded by the Singapore Maritime Institute (SMI) through grant number SMI-2015-AIMP-030. SL and the MERA authors acknowledge the Research Impact Grant from the University of Western Australia.

\section*{Appendix A}
\label{sec_appendix_fitted_gamma_jonswap}
\begin{figure}[ht]
    \centering
    {\includegraphics*[viewport= 128 307 450 518, width=0.49\linewidth]{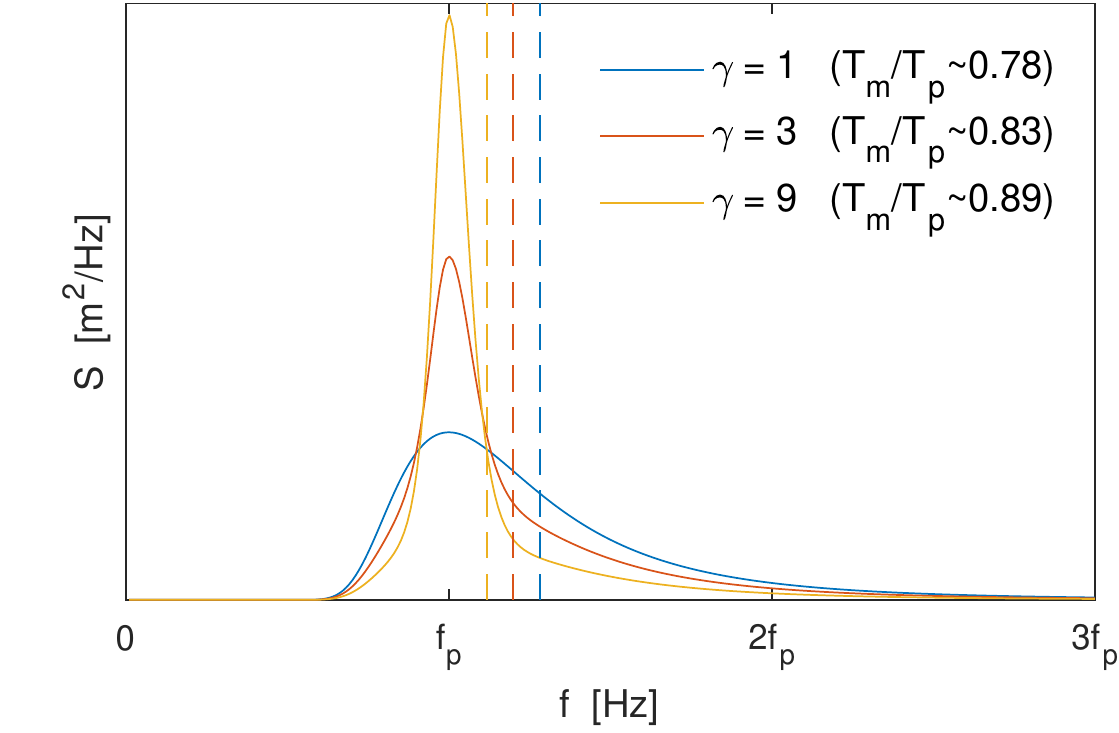}}
    \caption{JONSWAP spectra, each with the same peak period $T_p$ and the same significant wave height $H_s$, for different peak enhancement coefficients $\gamma$. The horizontal axis is expressed in terms of the peak frequency $f_p=\frac{1}{T_p}$. The dashed vertical lines indicate the mean frequency $f_m = \frac{1}{T_m}$, where $T_m$ is the mean period, with the line colour matching the corresponding spectrum. }
    \label{fig_JONSWAP}
\end{figure}

In the absence of the full spectral content in the Orkney hindcast dataset, we propose to approximate the unknown wave energy distribution by utilising JONSWAP spectral shapes with the hindcast bulk parameters of significant wave height $H_s$, peak period $T_p$ and mean period $T_m$. The mean period is matched approximately by choosing an appropriate value of the peak enhancement coefficient $\gamma$. We use a polynomial relation between the ratio $\frac{T_m}{T_p}$ and $\gamma$ given by
\begin{eqnarray}
\gamma &=& 13927 -68741 \frac{T_m}{T_p} + 127383 \Big(\frac{T_m}{T_p}\Big)^2 - 105074 \Big(\frac{T_m}{T_p}\Big)^3 + 32570 \Big(\frac{T_m}{T_p}\Big)^4 .      
\end{eqnarray}
The above function is derived through a least-squares fit, over the range $1 \le \gamma \le 10$, to the numerical data evaluated as per $T_m=\frac{m_0}{m_1}$, where $m_n = \int f^n S(f) \ \mathrm{d} f$ is the $n^{\mathrm{th}}$ spectral moment (with the frequency limits of the integrals set to 0 and $\frac{5}{T_p})$. Note that this relation is similar to \cite{ITTC}. Now, for each 3-hourly interval, we create a JONSWAP spectrum with inputs of $H_s$ and $T_p$ from the hindcast and input $\gamma$ estimated as above. These spectra $S(f)$ are then used in Equations \ref{eq_Pinc_finite} and \ref{eq_P}. Figure \ref{fig_JONSWAP} shows JONSWAP spectra for three different values of $\gamma$ illustrating its effect on the distribution of energy across frequency.

\begin{figure}[ht!]
\centering
{\includegraphics*[width=0.49\linewidth, viewport=150 308 400 518]{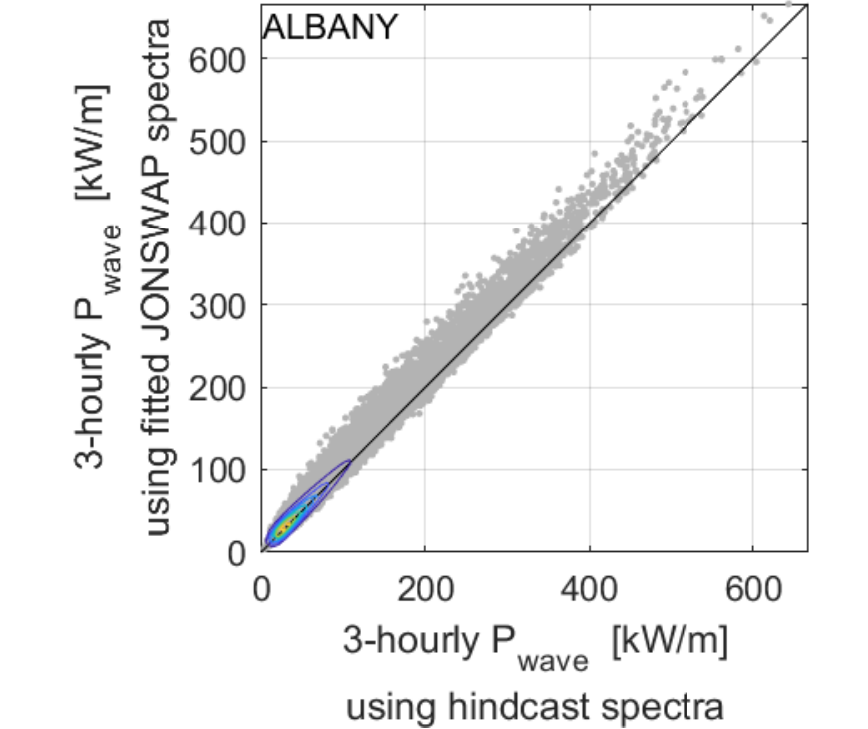}}
{\includegraphics*[width=0.49\linewidth, viewport=150 308 400 518]{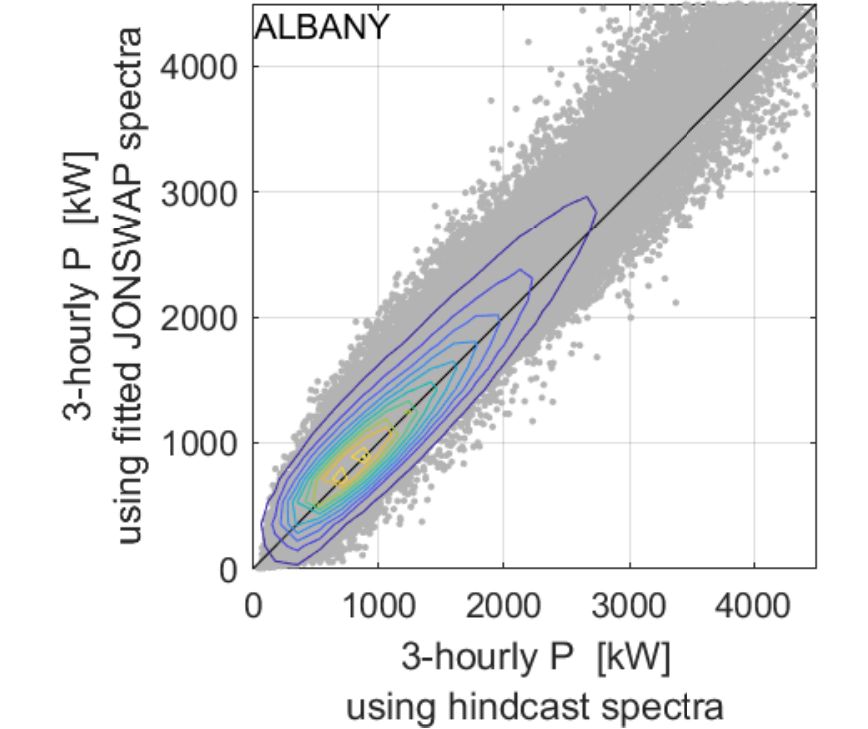}}
\caption{Comparison of calculated 3-hourly power values using the actual hindcast spectral shapes and using the fitted JONSWAP spectra (dot markers). Incident wave power density $P_{wave}$ is shown on the left and absorbed M4 power $P$ on the right. The contour lines indicate the probability of occurrence, with yellow lines corresponding to the more common values and green and blue lines corresponding to progressively less common instances. The black line represents 1:1 correspondence. Both plots are for Albany, where the full hindcast spectral information is available.}
\label{fig_P_wave_P_compare}
\end{figure}

For consistent power calculations at both locations, we also apply the above procedure to the Albany wave hindcast, for which the full spectra data is available. We can thus check how well the approximation works, which is displayed in Figure \ref{fig_P_wave_P_compare}. It is of course not possible to perfectly re-create the hindcast spectra by simply matching the three bulk parameters. This is manifested by the data spread in the plots. In particular, bi- and multi-modal sea-states cannot be accurately approximated by a single fitted JONSWAP shape. However, in the absence of the full spectral content, our approximation provides a sensible estimate of the energy distribution.  

\section*{Appendix B}
\label{sec_appendix_Pwave_histograms}
\begin{figure}[ht!]
\centering
{\includegraphics*[width=0.49\linewidth, viewport=120 303 447 520]{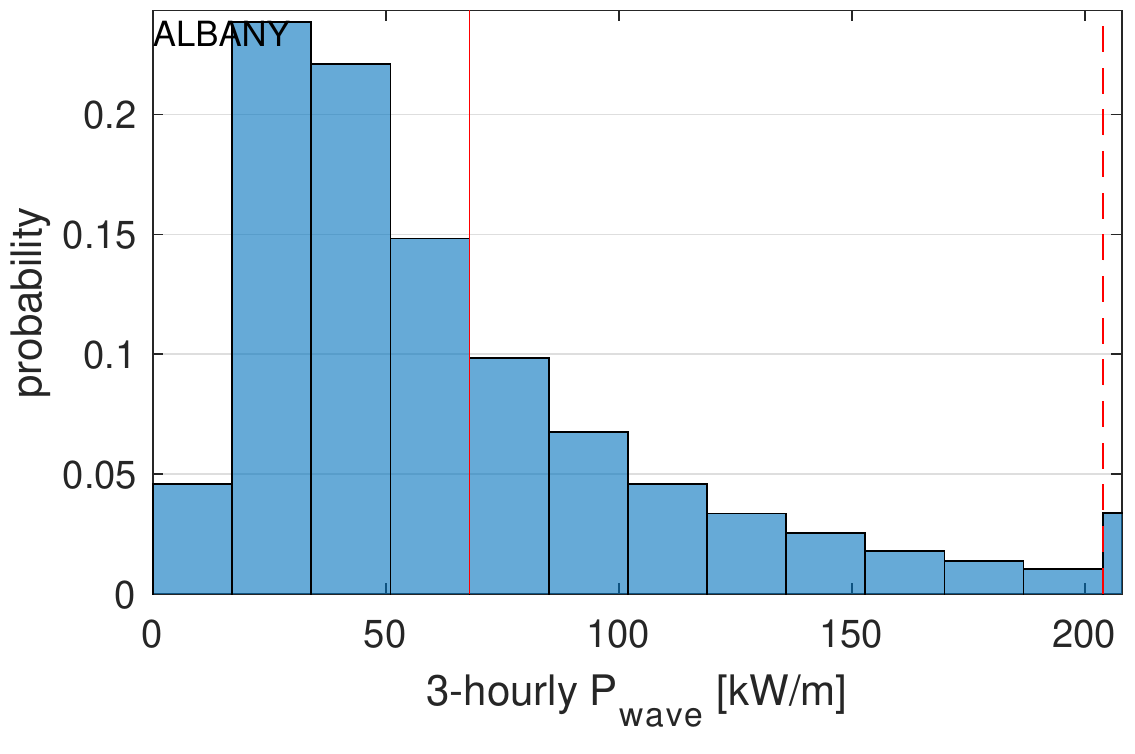}}
{\includegraphics*[width=0.49\linewidth, viewport=120 303 447 520]{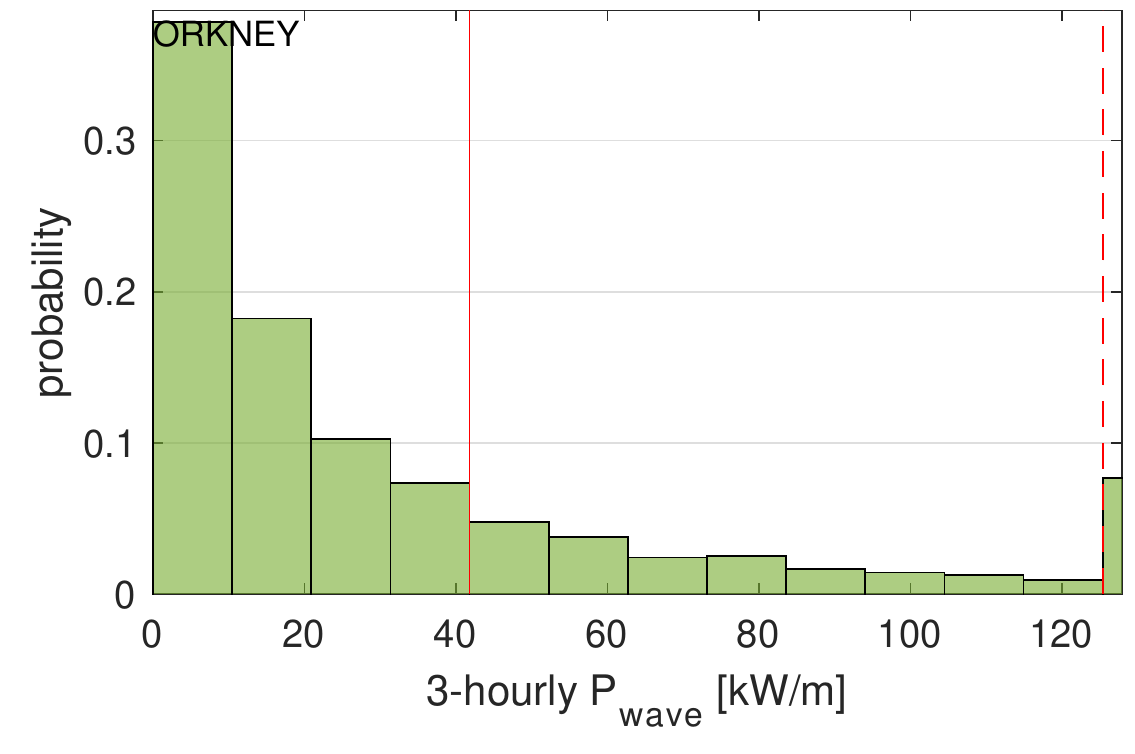}}
\caption{Histogram of average annual 3-hourly incident wave power density $P_{wave}$ for Albany on the left and for Orkney on the right. The long-term average incident wave power density $\bar{P}_{wave}$ is shown by the solid red line. The dashed red line represents three times this long-term value, with the last bin (only partially shown) containing all instances that exceed this.}
\label{fig_histogram_Pwave}
\end{figure}

Figure \ref{fig_histogram_Pwave} shows the distribution of 3-hourly incident wave power density at Albany during $1980 - 2017$ and at Orkney during $1958 - 2011$. Adopting the same plotting convention as in Figure \ref{fig_histogram_off_clip}, the long-term average at each location, denoted by $\bar{P}_{wave}$, is shown by the solid red line. The dashed red line denotes three times this value. The height of the last bin represents probability of occurrence of $P_{wave} > 3 \times \bar{P}_{wave}$.

\begin{figure}
\centering
{\includegraphics*[width=1\linewidth, viewport=22 280 540 570]{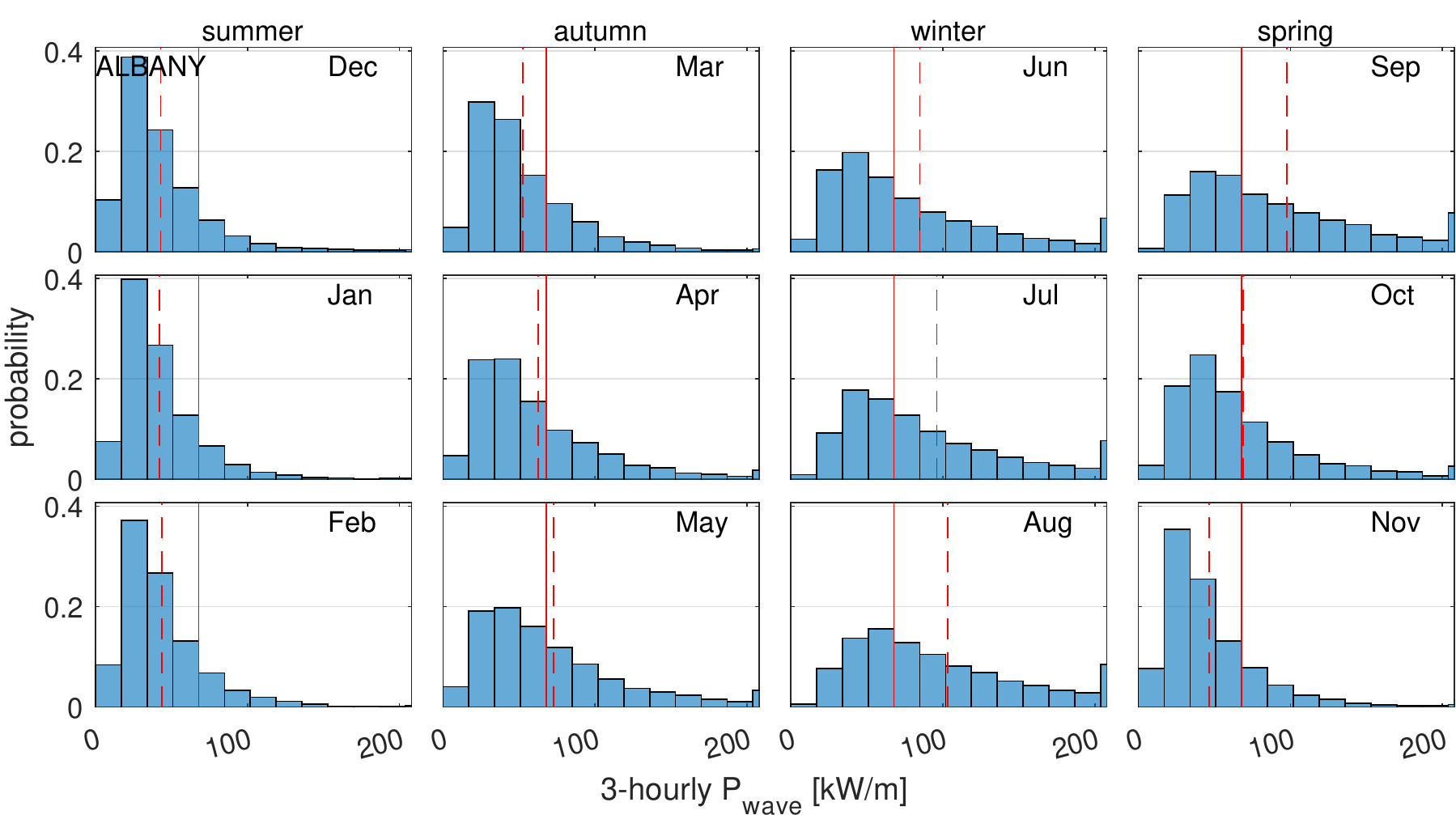}}
{\includegraphics*[width=1\linewidth, viewport=22 280 540 570]{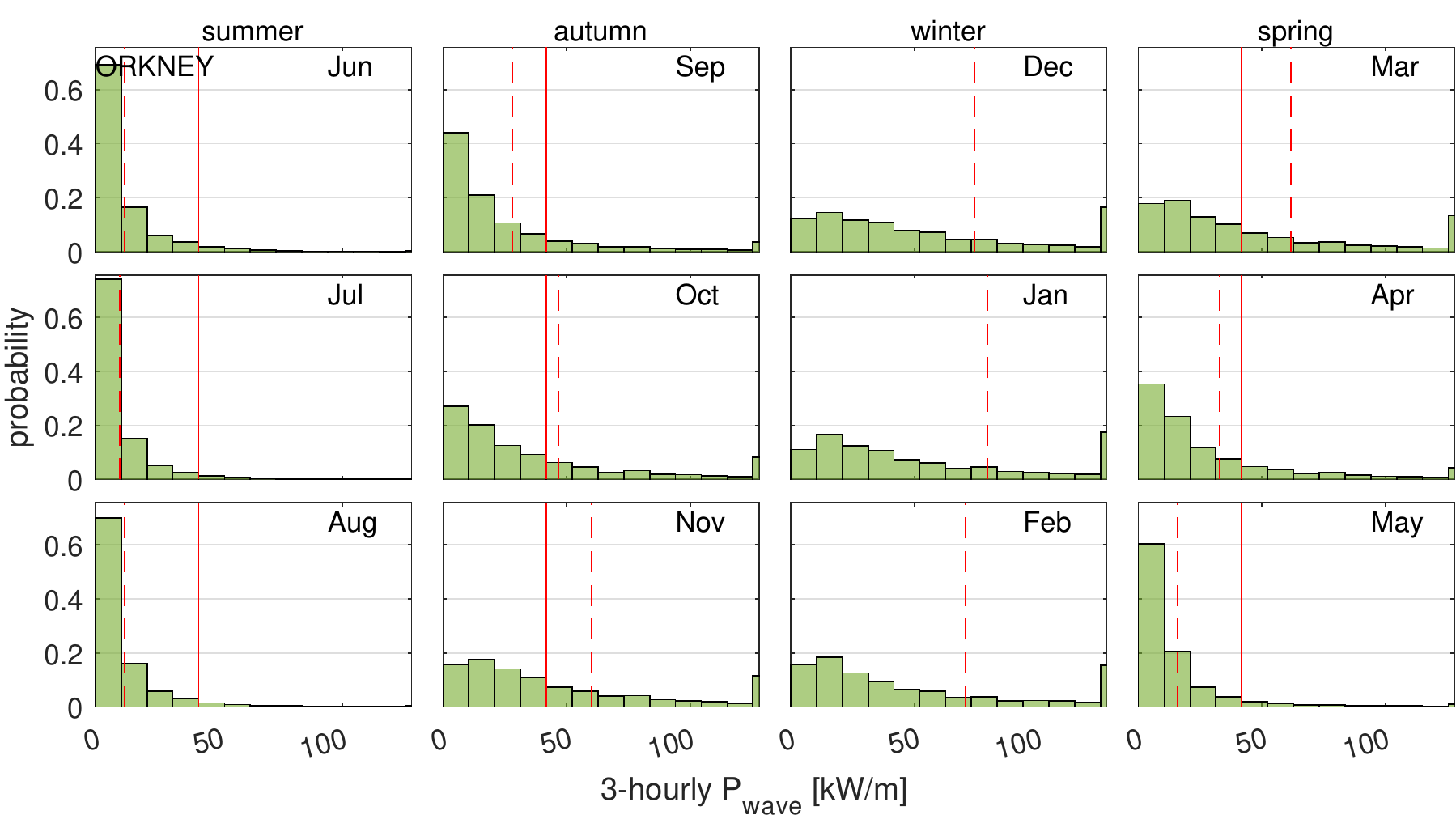}}
\caption{Histogram of 3-hourly incident wave power density $P_{wave}$ split into 12 months for Albany at the top and for Orkney at the bottom. The seasons at the two locations have been aligned, such that the columns (of subplots) from left to right correspond to summer, autumn, winter and spring. The long-term average incident wave power density $\bar{P}_{wave}$ is shown by the solid red line, while the dashed red line represents the long-term average for each month. Note that the last bin in each histogram  (only partially shown) contains all instances that exceed $3 \times \bar{P}_{wave}$.}
\label{fig_histogram_Pwave_monthly}
\end{figure}

Figure \ref{fig_histogram_Pwave_monthly} presents the 3-hourly incident wave power density distributions for each month. The long-term average $\bar{P}_{wave}$ is shown by the red solid line, while the long-term average for each month is shown by dash-dotted lines. The height of the last bin represents probability of occurrence of $P_{wave} > 3 \times \bar{P}_{wave}$.

\bibliographystyle{elsarticle-harv}
\bibliography{ref}


\label{lastpage}

\end{document}